\documentclass[10pt,conference]{IEEEtran}
\IEEEoverridecommandlockouts
\usepackage{cite}
\usepackage{amsmath,amssymb,amsfonts}
\usepackage{algorithmic}
\usepackage{graphicx}
\usepackage{textcomp}
\usepackage{xcolor}
\usepackage[hyphens]{url}
\usepackage{enumitem}
\usepackage{fancyhdr}
\usepackage{hyperref}
\usepackage{mathptmx} 
\usepackage{multirow}
\usepackage{fancyhdr}
\usepackage[normalem]{ulem}
\usepackage{optidef}

\usepackage[T1]{fontenc}
\usepackage{flushend}
\usepackage{balance}
\usepackage{booktabs}
\usepackage{epsfig}
\usepackage{lipsum}
\usepackage[caption=false,font=footnotesize]{subfig}
\usepackage{color}

\usepackage[normalem]{ulem}
\usepackage{bigstrut}
\usepackage{multirow}
\usepackage{ulem}
\usepackage{listings}
\usepackage{eso-pic}
\usepackage{tikz}
\usepackage{xspace}
\usepackage{linegoal}
\usepackage{mathtools}
\usepackage{soul}
\usepackage[leftcaption]{sidecap}
\def\BibTeX{{\rm B\kern-.05em{\sc i\kern-.025em b}\kern-.08em
    T\kern-.1667em\lower.7ex\hbox{E}\kern-.125emX}}
\begin{document}

\makeatletter 
\newcommand{\linebreakand}{%
  \end{@IEEEauthorhalign}
  \hfill\mbox{}\par
  \mbox{}\hfill\begin{@IEEEauthorhalign}
}
\makeatother 
 \newcommand{\todo}[1]{\textcolor{red}{#1}}
    \newcommand{\cortexm}{\mbox{Cortex-M}}
    \newcommand{\zoix}{Synopsys Z01X\mbox{\texttrademark~\cite{z01x}}}
\title{Characterizing Soft-Error Resiliency in Arm's Ethos-U55 Embedded Machine Learning Accelerator}

\author{
    \IEEEauthorblockN{Abhishek Tyagi}
    \IEEEauthorblockA{\textit{Department of Computer Science} \\
    \textit{University of Rochester}\\
    Rochester, NY, USA \\
    atyagi2@ur.rochester.edu}
\and
    \IEEEauthorblockN{Reiley Jeyapaul}
    \IEEEauthorblockA{\textit{Reliability, Availability, and Serviceability (RAS)}\\
    \textit{AMD}\\
    Austin, TX, USA \\
    reiley.jeyapaul@ieee.org}
\and
    \IEEEauthorblockN{Chuteng Zhou}
    \IEEEauthorblockA{\textit{Central Technology}\\
    \textit{ARM Inc}\\
    Austin, TX, USA \\
    chu.zhou@arm.com}
\and 
\linebreakand
    \IEEEauthorblockN{Paul Whatmough}
    \IEEEauthorblockA{\textit{AI Research}\\
    \textit{Qualcomm}\\
    Boston, MA, USA \\
    pwhatmou@qti.qualcomm.com}
\and
    \IEEEauthorblockN{Yuhao Zhu}
    \IEEEauthorblockA{\textit{Department of Computer Science}\\
    \textit{University of Rochester}\\
    Rochester, NY, USA \\
    yzhu@rochester.edu}
}

\maketitle
\thispagestyle{plain}
\pagestyle{plain}





\newcommand{\website}[1]{{\tt #1}}
\newcommand{\program}[1]{{\tt #1}}
\newcommand{\benchmark}[1]{{\it #1}}
\newcommand{\fixme}[1]{{\textcolor{red}{\textit{#1}}}}

\newcommand*\circled[2]{\tikz[baseline=(char.base)]{
            \node[shape=circle,fill=black,inner sep=1pt] (char) {\textcolor{#1}{{\footnotesize #2}}};}}

\ifx\figurename\undefined \def\figurename{Figure}\fi
\renewcommand{\figurename}{Fig.}
\renewcommand{\paragraph}[1]{\textbf{#1} }
\newcommand{\figline}{{\vspace*{.05in}\hline}}

\newcommand{\Sect}[1]{Sec.~\ref{#1}}
\newcommand{\Fig}[1]{Fig.~\ref{#1}}
\newcommand{\Tbl}[1]{Tbl.~\ref{#1}}
\newcommand{\Eqn}[1]{Eqn.~\ref{#1}}
\newcommand{\Apx}[1]{Apdx.~\ref{#1}}
\newcommand{\Alg}[1]{Algo.~\ref{#1}}

\newcommand{\specialcell}[2][c]{\begin{tabular}[#1]{@{}c@{}}#2\end{tabular}}
\newcommand{\note}[1]{\textcolor{red}{#1}}

\newcommand{\proj}{\textsc{Cicero}\xspace}
\newcommand{\algo}{\textsc{SpaRW}\xspace}
\newcommand{\mode}[1]{\underline{\textsc{#1}}\xspace}
\newcommand{\sys}[1]{\underline{\textsc{#1}}}

\newcommand{\no}[1]{#1}
\renewcommand{\no}[1]{}
\newcommand{\RNum}[1]{\uppercase\expandafter{\romannumeral #1\relax}}

\def\cG{{\mathcal{G}}}
\def\cF{{\mathcal{F}}}
\def\cI{{\mathcal{I}}}
\def\cN{{\mathcal{N}}}
\def\bh{{\mathbf{h}}}
\def\bp{{\mathbf{p}}}


\begin{abstract}
As Neural Processing Units (NPU) or accelerators are increasingly deployed in a variety of applications including safety critical applications such as autonomous vehicle, and medical imaging, it is critical to understand the fault-tolerance nature of the NPUs.
We present a reliability study of Arm's Ethos-U55, an important industrial-scale NPU being utilised in embedded and IoT applications. 
We perform large scale RTL-level fault injections to characterize Ethos-U55 against the Automotive Safety Integrity Level D (ASIL-D) resiliency standard commonly used for safety-critical applications such as autonomous vehicles. We show that, under soft errors, all four configurations of the NPU fall short of the required level of resiliency for a variety of neural networks running on the NPU. 
 
We show that it is possible to meet the ASIL-D level resiliency without resorting to conventional strategies like Dual Core Lock Step (DCLS) that has an area overhead of 100\%.
We achieve so through \textit{selective protection}, where hardware structures are selectively protected (e.g., duplicated, hardened) based on their sensitivity to soft errors and their silicon areas.
To identify the optimal configuration that minimizes the area overhead while meeting the ASIL-D standard, the main challenge is the large search space associated with the time-consuming RTL simulation.
To address this challenge, we present a statistical analysis tool that is validated against Arm silicon and that allows us to quickly navigate hundreds of billions of fault sites without exhaustive RTL fault
injections.
We show that by carefully duplicating a small fraction of the functional blocks and hardening the Flops in other blocks meets the ASIL-D safety standard while introducing an area overhead of only 38\%.



\end{abstract}
\section{Introduction}
Machine learning accelerators, especially those that target Deep Neural Networks (DNNs), are increasingly used in safety-critical applications, such as autonomous vehicles~\cite{tesla, gan2022braum, gan2020ptolemy} and medical devices~\cite{gertz2020applications}. Ensuring reliable and resilient operations have become essential~\cite{zhu2017cognitive}.
Among all sources of vulnerabilities, we focus on soft errors~\cite{papadimitriou2021demystifying,sridharan2009eliminating,seifert2006radiation}, which are transient faults induced by radiation or other external factors (e.g., voltage droops) that can compromise the integrity of data and computations within an NPU.
This paper focuses on Arm's Ethos-U55~\cite{u55} microNPU, a commercial DNN accelerator used primarily for embedded applications. We provide a thorough characterization of U55's resiliency against soft errors and evaluate how U55's resiliency is impacted by a number of commonly used soft-error mitigation techniques.

Using the RTL of U55, we perform a large-scale fault injection campaign (Sec~\ref{sec:se_char}). We show that U55, across a range of hardware configurations and DNNs, shows a Silent Data Corruption (SDC) rate lower than $0.1 \times 10^{-15}$ per inference. While exceedingly low and indeed lower than (i.e., satisfies) the Automotive Safety Integrity Level (ASIL) B and C standards, the SDC rate still violates the ASIL-D standard, the most strict form of ASIL. The SDC rate, perhaps unsurprisingly, increases with the scale of the NPU (e.g., MAC array/on-chip SRAM sizes).

We then dive deeper into individual functional blocks in the U55 NPU. We show that different functional blocks in the NPU (e.g., MAC array vs. DMA vs. control block) have inherently different sensitivity toward soft errors: generally the units responsible for managing dataflow and for decoding weights from memory, when experiencing a soft error, could lead to a higher rate of overall system SDC than other hardware structures. Critically, this sensitivity pattern holds under different process nodes but changes significantly depending on whether faults in logic elements are considered.

We then characterize how U55's soft-error resiliency can be improved by common, existing soft-error protection/mitigation techniques (Sec~\ref{sec:ext_se}).
This is an important study because all protection techniques, such as modular redundancy~\cite{teifel2008self,giterman2016area} or flop hardening~\cite{calin1996upset, jahinuzzaman2009soft,li2017quatro,jagannathan2011single}, introduce area overhead \footnote{They will introduce power overhead too, but we are not allowed to share detailed power results.}.
However, we find that different function blocks in U55 have different area-vs-resiliency trade-offs.
For instance, the control unit tends to be small but is sensitive to soft errors.
Therefore, there exists an optimal protection strategy given an area budget, which is an important figure of merit in embedded applications as U55 is commonly used.

To characterize the area-vs-resiliency trade-off of U55, we present an internal statistical analysis tool that is validated against Arm silicon and that allows us to quickly navigate hundreds of billions of fault sites without exhaustive RTL fault injections.
We show that in order to meet the most stringent ASIL-D standard, some form of modular redundancy must be introduced. However, one does not have to duplicate all the function blocks. In particular, Ethos-U55 meets ASIL-D standard when only Traversal Unit (TSU) and Weight Decoder (WD) blocks are duplicated and DMA and MAC Unit blocks have their FFs hardened.

In summary, this paper makes the following contributions:
\begin{itemize}
    \item To the best of our knowledge, this is the first large-scale resiliency characterization of a commercial NPU based on RTL fault injections. See Sec\mbox{~\ref{sec:related}} for comparison with prior works in commercial accelerator reliability analysis.
    \item We report the soft error resiliency of all the key functional blocks of Ethos-U55 NPU; these blocks are representative as they are found in common ML inference processors in the industry. Such reliability analysis helps us understand, at a per functional block level, the overhead-vs-resiliency trade-off of various protection mechanisms and how they affect the overall reliability of the IP.
    \item We also show that when searching for soft-error detection strategies to meet the highest safety standards under silicon area constraints, it is in the designer's interest to look at a mixture of detection schemes rather than choosing one scheme for the entire IP.
    \item We describe a fast and faithful resiliency characterization methodology used inside Arm. The methodology combines functional block level RTL fault injection (using Synopsys Z01X~\mbox{\cite{z01x}}) and (RTL-validated) statistical fault analysis (Thales~\mbox{\cite{tyagi2022thales}}).
\end{itemize}

\section{Background}
We first describe the scope of our work (Sec~\ref{sens:scope}). We then describe the basics of soft-errors (Sec~\ref{sec:softerror}). We describe in detail the architecture and use cases of Arm's Ethos-U55 (Sec~\ref{sec:u55}). We end the section by discussing the existing methods of soft-error resilience (Sec~\ref{sens:dmr}).

\begin{figure}[t]
	\centering
	\includegraphics[width=\columnwidth,height=0.8\columnwidth]{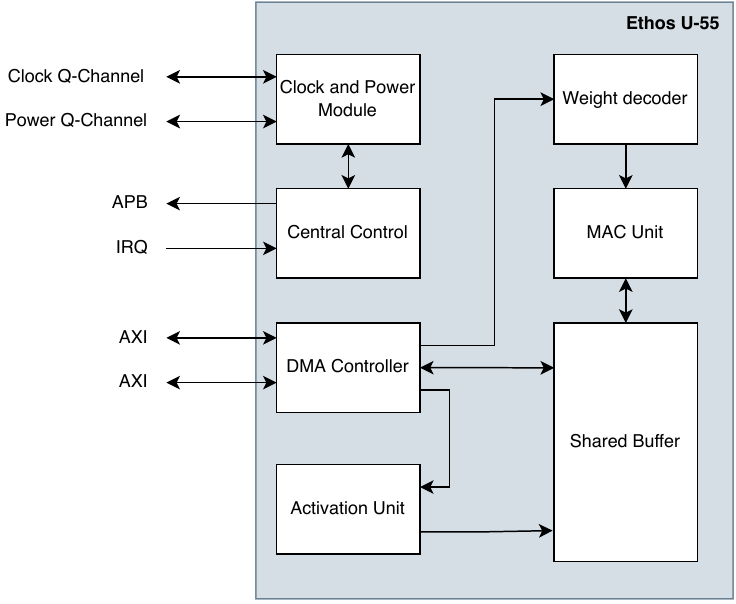}
	\caption{Ethos-U55 functional blocks diagram~\cite{u55_block_diag}}
\label{fig:npu}

\end{figure}

\subsection{Scope and Assumptions}
\label{sens:scope}
We are interested in characterizing the soft-error reliability of Ethos-U55~\cite{u55}. While transient soft-errors can occur anywhere on the chip~\cite{naseer2007critical,jahinuzzaman2009soft,degalahal2003analyzing,lantz1996soft,de2015evaluation}, they are most damaging to logic structures and Flip-Flops (FF); other storage structures are usually protected by error correction codes~\cite{slayman2005cache,chen1984error}. 
In this work, we use FIT rate and SDC rate as metrics to quantify the reliability of the NPU. In the context of DNN accelerators, an SDC is an inference mis-prediction~\cite{chen2020tensorfi}, whereas FIT rate not only considers SDCs but crashes as well.

\subsection{Soft-Errors}
\label{sec:softerror}
The Single Event Effects (SEEs) encompass both Single Event Transients (SETs) and Single Event Upsets (SEUs). A SET occurs as a voltage glitch at the output of a combinational gate when an incident particle deposits adequate charge in the gate's sensitive region. Subsequently, the SETs can propagate to sequential cells and induce a change in the stored logic value, leading to a soft error or SEU. Alternatively, soft errors may result from energetic particles directly impacting sequential logic components like flip-flops and latches.

\subsection{ Ethos-U55 Overview}
\label{sec:u55}
Ethos-U55 is Arm's first microNPU designed for the embedded market and meets the requirements for performance with low area and power giving 90\% energy reduction with up to 480x performance increase as compared to Cortex-M series alone. U55 is designed to operate while coupled to \cortexm~\cite{m0,m1} series processors, which act as controllers.  

U55 is being used widely in the market by companies such as Alif Semiconductors for their Ensemble Series of IP. Their E1, E3, E5, and E7 series~\cite{semiconductorintroducing} uses U55 for applications such as wearables, security camera systems, medical devices, and retail applications. NXP has also integrated U55 and U65~\cite{u65} with their i.MX~\cite{nxp} series of processors to be used in systems such as driver monitoring systems in the automotive sector. With U55 being utilized in safety-critical applications, it becomes important to characterize the inherent reliability of the design for meeting stringent safety requirements.


Fig~\ref{fig:npu} showcases the various functional blocks comprising the IP, whose functionality is described below: 
\begin{itemize}
	\item \textbf{Direct Memory Access (DMA) Controller}: manages the movement of data from external memory to on-chip memory.
	\item \textbf{Central Controller (CC)}: is responsible for managing the distribution of tasks to all the units in the NPU. We divide the CC in two parts: 
	\begin{itemize}
		\item{\textbf{Traversal Unit (TSU)}}: manages the dataflow to and from the MAC unit to maintain correct execution
		\item{\textbf{Register File (REG)}}: stores configuration values in CC.
	\end{itemize}
	\item \textbf{Weight Decoder (WD)}: reads the weights from either on-chip RAM or from an internal buffer and dispatches weights to the MAC array.
	\item \textbf{MAC Array}: carries out the Multiply and Accumulate operations on the input and weights.
	\item \textbf{Activation Unit (AO)}: receives the output feature map from the MAC array and can apply either activation functions or add bias to the read values.
	\item \textbf{Shared Buffer}: is used to store the intermediate output feature maps, input activations, and/or weights. We assume that memory structures (such as SRAM, DRAM) are protected using either ECC and/or parity~\cite{mahmoud2021optimizing}.
\end{itemize}

\subsection{Existing Soft-Error Resilient Approaches}
\label{sens:dmr}
\paragraph{Dual Modular Redundancy (DMR)} duplicates a functional block executing a program and comparing the two sets of outputs. If the checking logic detects any mismatch between the outputs, OS can send a request a re-execution of the program or employ recovery mechanisms. DMR is an effective detection strategy against bit-flips incurred by direct charge injection in a FF and also for latching a SET due to an error injection in combinational logic.

\paragraph{Flop Hardening} modifies a FF such that a bit-flip is significantly less likely to take place. While a hardened FF will make it difficult for a particle strike to flip the bit stored in the FF, it would not prevent the FF from latching onto a SET that reaches the input of the FF~\cite{seifert2012soft}. The Dual Interlocked Storage Cell (DICE) is a widely utilized custom rad-hard flip-flop design~\cite{calin1996upset}. An alternative approach, Quatro, based on Cascode Voltage Switch Logic (CVSL), has been proposed to achieve better performance at high LET values~\cite{jahinuzzaman2009soft,li2017quatro,jagannathan2011single}. Some custom flip-flop designs have also addressed Single Event Transients (SETs). For example, an improved DICE implementation with integrated tunable delay elements for SET filtering was suggested~\cite{knudsen2006area}. 



\section{Ethos-U55 Soft Error Characterization}
\label{sec:se_char}
We first describe our methodology for obtaining the characterization data (\Sect{sec:fi_setup}). We then describe how the SoC FIT rate is translated to NPUs SDC per inference (\Sect{sec:sdc_fit}). We then put forward the resiliency data for Ethos-U55 for various configurations and applications (\Sect{sec:sdc_overall}). We then describe in detail various factors constituting the resiliency behavior of Ethos-U55 (\Sect{sec:fb_fac}). And finally, we end the section by discussing the absence of a correlation between area of a functional block and its inherent resiliency (\Sect{sec:sdc_area}).
\subsection{Fault Injection Setup}
\label{sec:fi_setup}

\begin{table}[t]
	\centering
	\caption{Workloads used for soft-error resiliency characterization. ASR standards for Automatic Speech Recognition.}
	\renewcommand*{\arraystretch}{1}
	\renewcommand*{\tabcolsep}{4pt}
	\resizebox{\columnwidth}{!}
	{
		\begin{tabular}{cccc}
			\toprule[0.15em]
			\textbf{Category} & \textbf{Total Parameters} & \textbf{Network}   & \textbf{Dataset}  \\
			\midrule[0.05em]
			Classification      & $1.1\times 10^6$ & CifarNet~\cite{hosang2015taking}     &CIFAR-10~\cite{krizhevsky2009learning}\\
			Classification     & $11.2\times 10^6$  & ResNet-18~\cite{he2016deep}   & ImageNet~\cite{deng2009imagenet}\\ 
			ASR & $23\times 10^6$ & Wav2Letter~\cite{collobert2016wav2letter}   & LibriSpeech~\cite{panayotov2015librispeech}\\
			\bottomrule[0.15em]
		\end{tabular}
	}
	\label{tbl:prob}
\end{table}

We use RTL fault injection to obtain precise soft-error resiliency data. We use Synopsys Z01X\texttrademark~\cite{z01x}, which is an industrial-scale RTL fault injection tool, to pick hardware fault sites, represented as \mbox{$<\mathbf{Cycle, FF, BP}>$}, to flip.

For a single run, the tool picks a single fault-site to flip and performs the RTL simulation that runs the application to the end. For each application, we inject over 2 million faults into the Arms Ethos U55 RTL and run the RTL simulations. This ensures that the resiliency data has less than 1\% of error margin with a 99\% confidence interval per application.

\Tbl{tbl:prob} lists the applications we use to evaluate the reliability of Arms Ethos U55. We choose applications that are utilized in safety-critical scenarios and vary in their sizes, as the size of a network has previously been shown to affect the resiliency of neural networks\mbox{~\cite{ibrahim2020soft}}. CifarNet\mbox{~\cite{hosang2015taking}} is a widely used neural network in embedded autonomous platforms\mbox{~\cite{kocic2019end}}. ResNet-18\mbox{~\cite{he2016deep}} is utilized as the backbone of the majority of object detection networks deployed in autonomous vehicles (traditional object detection networks are not supported on U55\mbox{~\cite{mlzoo}}. We also use Wav2Letter\mbox{~\cite{collobert2016wav2letter}}, an automatic speech recognition (ASR) network by Meta. ASR has been utilized in safety-critical systems such as aviation to improve flight efficiency and air traffic control to improve communications\mbox{~\cite{ahrenhold2023validating,kleinert2021automated}}.


\subsection{Translating SoC FIT Rate to NPU SDC per Inference}
\label{sec:sdc_fit}
Synopsys Z01X\texttrademark  compares the output of a fault-injected run with a faultless run and flags an error on output mismatch.
For our applications, we consider an SDC to take place when there is a top-1 label mismatch for the image classification tasks and a decrease in word error rate (WER) for the ASR task. 
We note that using the per-inference misprediction approach, while applies to image classification and ASR tasks this paper focuses on, may not apply to all ML tasks; the notion of SDC, indeed, must be defined on a per-task basis, because different tasks have different task-level masking.
For LLMs\mbox{~\cite{devlin2018bert,touvron2023llama}} and generative AI tasks\mbox{~\cite{ramesh2021zero,oppenlaender2022creativity}}, it is still an open question as to how SDCs should be defined.


For applications such as self-driving cars, the overall FIT rate of the chipset should be less than 10 (failures) in 1 billion hours of operation to meet ASIL-D standards for critical components such as airbags and antilock braking~\cite{synopsys}.
Other ASIL standards are more relaxed. Specifically, for ASIL-B and ASIL-C standards (enforced on brake lights and active suspension~\cite{synopsys}) the FIT rate should be less than 100 (failures) per 1 billion hours of operation.

Since NPU is just a fraction of an SoC, the NPU's FIT rate requirement should just be a fraction of that of the SoC.
The fraction equals the area of the NPU with respect to the entire SoC, as described by Li et al.~\cite{li2017understanding} and Fidelity~\cite{he2020fidelity}. For an SoC such as Tesla FSD Chip~\cite{tesla}, an Ethos-U55 will occupy a fraction of the area depending on the MAC configuration (0.12 for MAC-32, 0.14 for MAC-64, 0.17 for MAC-128 and 0.27 for MAC-256). Based on the fraction of area, the FIT requirements for each MAC configuration become 0.12, 0.14, 0.17, and 0.27 failures per $10^9$ hours for MAC-32, MAC-64, MAC-128, and MAC-256 respectively.

We use Synopsys Z01X\texttrademark~to perform RTL-level fault injections, which provides the relative FIT rate \textit{assuming} a fault has occurred. Based on the raw FIT rate data of flip flops~\cite{cao2019alpha} and the inference time of a DNN, we can then calculate the absolute FIT rate of the NPU.
Using a conservative inference time of 0.3 ms (which accounts for the slowest running application in our experiments), 
we estimate that the required FIT rate has to be less than $0.1 \times 10^{-15}$, $0.12 \times 10^{-15}$, $0.15 \times 10^{-15}$ and $0.23 \times 10^{-15}$ to be comparable with ASIL-D standards for MAC-32, MAC-64, MAC-128, and MAC-256 configurations respectively.

As mentioned previously, while calculating FIT rate, SDC and crashes both are considered. 
We show in \Sect{sec:sdc_overall}, that even while considering just the SDCs for Ethos-U55, its resiliency falls short of ASIL-D standards. Moreover, crashes are easier to detect than SDCs and do not require the same amount of overhead as SDC detection and protection. Due to these reasons, we do not consider crashes in this work and hence can use the FIT rate calculated above as the required SDC rate per inference to meet the ASIL-D standards.
\subsection{How Resilient is Ethos-U55 to Soft Errors?}
\label{sec:sdc_overall}
\begin{figure}[t]
	\centering
	\includegraphics[width=\columnwidth]{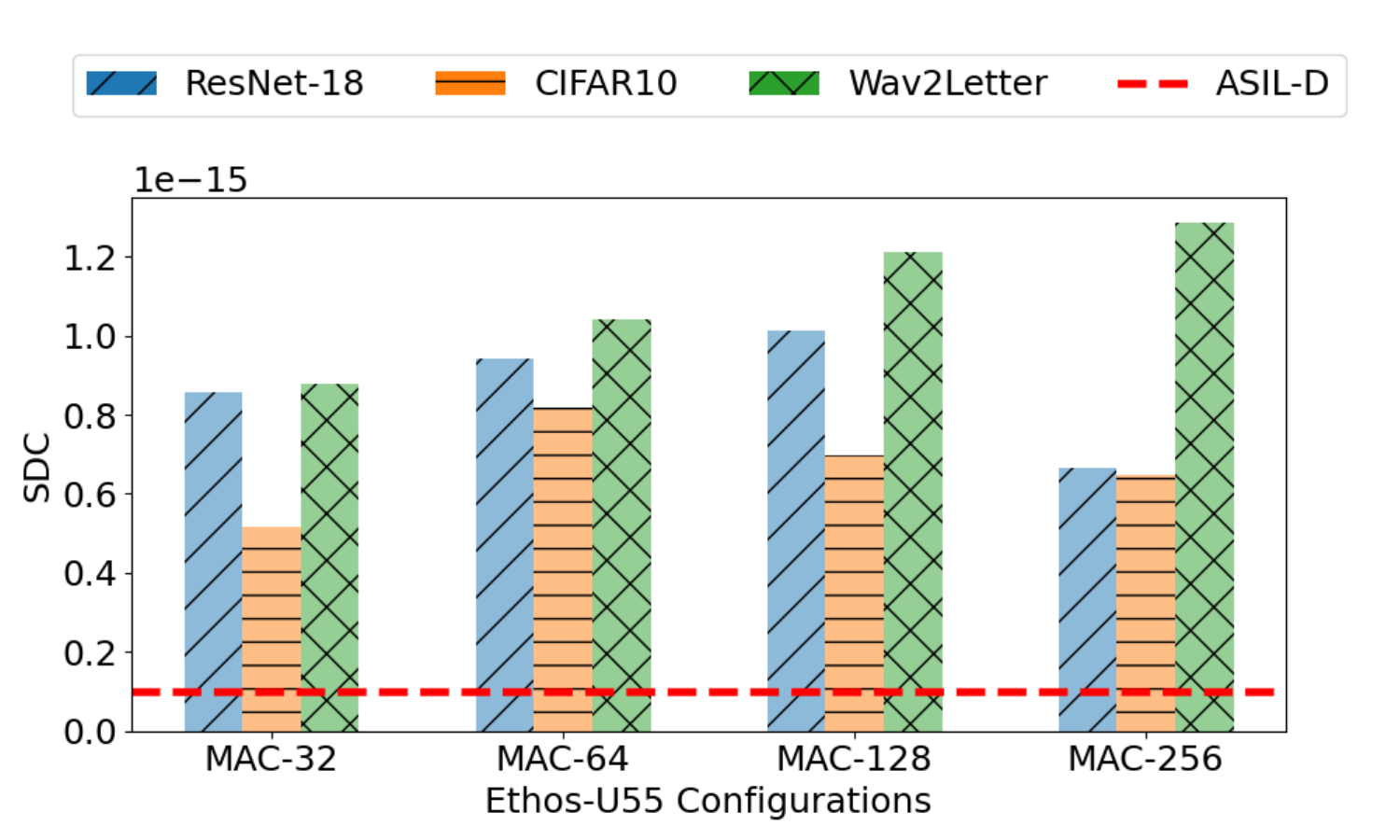}
	\caption{SDC Rate of Ethos-U55 while running ResNet-18, CifarNet, and Wav2Letter at TSMC 16nm technology node.}
\label{fig:sdc_overall_16nm}
\end{figure}


Fig~\ref{fig:sdc_overall_16nm} shows that the SDC rate of the NPU varies significantly with the application it is running as well as the underlying hardware configuration. CifarNet~\cite{hosang2015taking} consistently performs best on the resiliency aspects on all the four Ethos-U55 configurations whereas, Wave2Letter~\cite{collobert2016wav2letter} is the worst performing application on all the hardware configurations. For the given set of applications, MAC-32 configuration is most resilient to soft-errors.

More interestingly, as per our experiments, Ethos-U55 does not to meet the ASIL-D standards as the reported SDC rate (or the FIT) is $\geq 0.1\times 10^-15$ for all configurations. Therefore, it becomes important to understand what factors affect the resiliency of Ethos-U55. We dissect the NPU and look at the contribution of each individual functional block in the NPU to the overall resiliency of Ethos-U55, for all the different MAC configurations, running the given applications, on a chip fabricated in possible different technology nodes.

\subsection{Factors Shaping Functional Block Resilience}
\label{sec:fb_fac}
\subsubsection{Sensitivity to MAC Sizes}
\label{sens:mac}
\begin{figure}[t]
  \centering
  \includegraphics[width=\columnwidth]{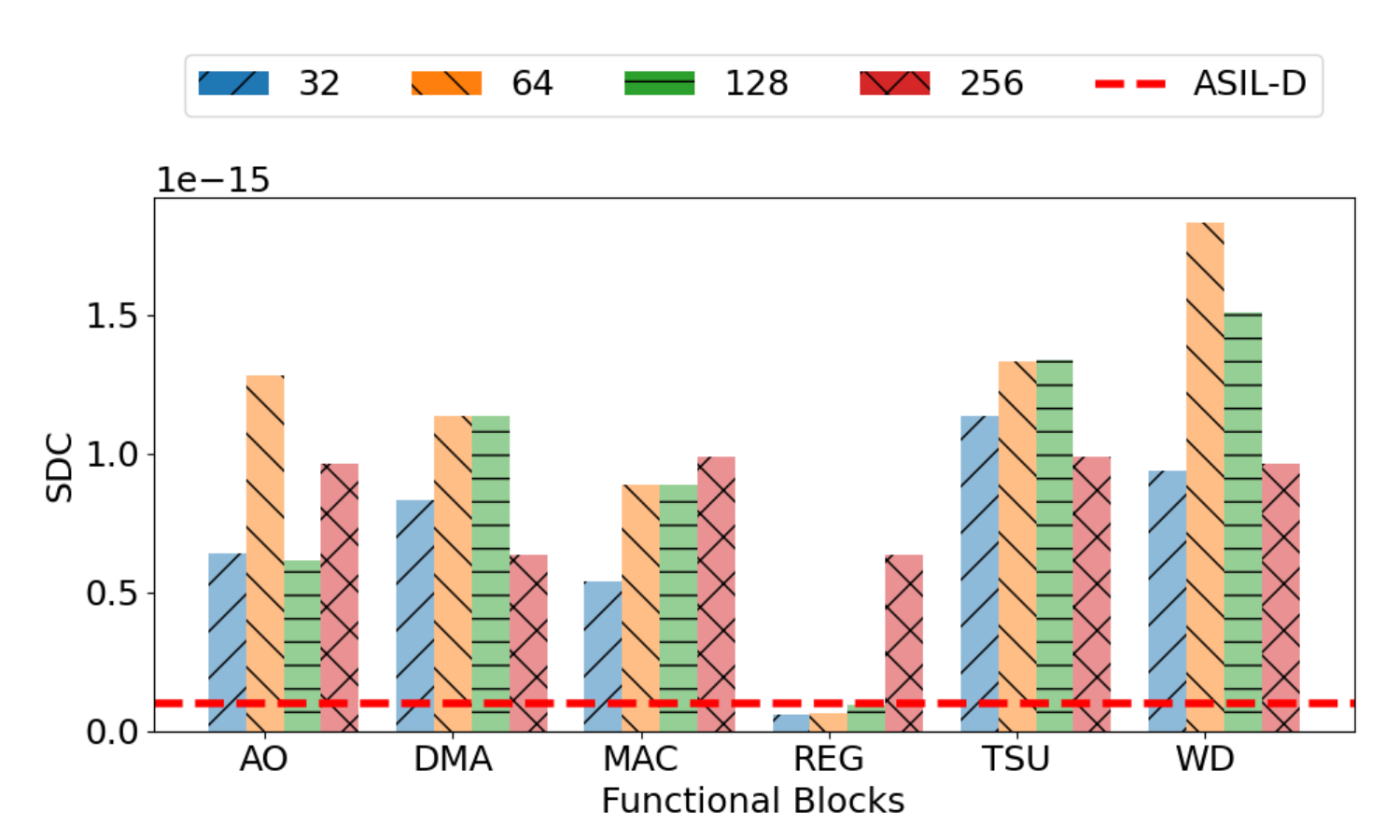}
  \caption{Functional block SDC contribution for different configurations of Arm Ethos-U55.}
  \label{fig:sens_mac}
\end{figure}
The resiliency of NPU functional blocks is sensitive to MAC array size which dictates how many multiply-and-accumulate computations can take place at any point in time. It also dictates the manner in which any large computation is broken down into smaller tasks, which affects the reuse of weights and/or activations. Intuitively, this changes the number and/or position of the faulty neuron in the neural network, eventually resulting in a variation in the MAC SDC rate.
Fig~\ref{fig:sens_mac} shows the variation in SDC contribution of each functional block running CifarNet on four possible configurations of Arm Ethos-U55, proving our intuition correct. 

In addition, as shown in Fig~\ref{fig:sens_mac}, the sensitivity of the SDC rate of the MAC unit to changes in MAC fabric size is logical, but it is surprising that other functional blocks also exhibit sensitivity to this variation. We see such a behavior because a change in MAC fabric size changes the task execution chunk size. To adapt to the new chunk size, all other functional blocks in the IP have to modify their execution flow which changes the block ultimately affecting the SDC rate of the block. 


\subsubsection{Sensitivity to Applications}
\label{sens:app}
Applications running footprint on the NPU impact the resiliency of a functional block within the IP. Each of the applications have a unique utilization footprint on each of the functional blocks which leads to varying performance on the underlying hardware.
\begin{figure}[t]
  \centering
  \includegraphics[width=\columnwidth]{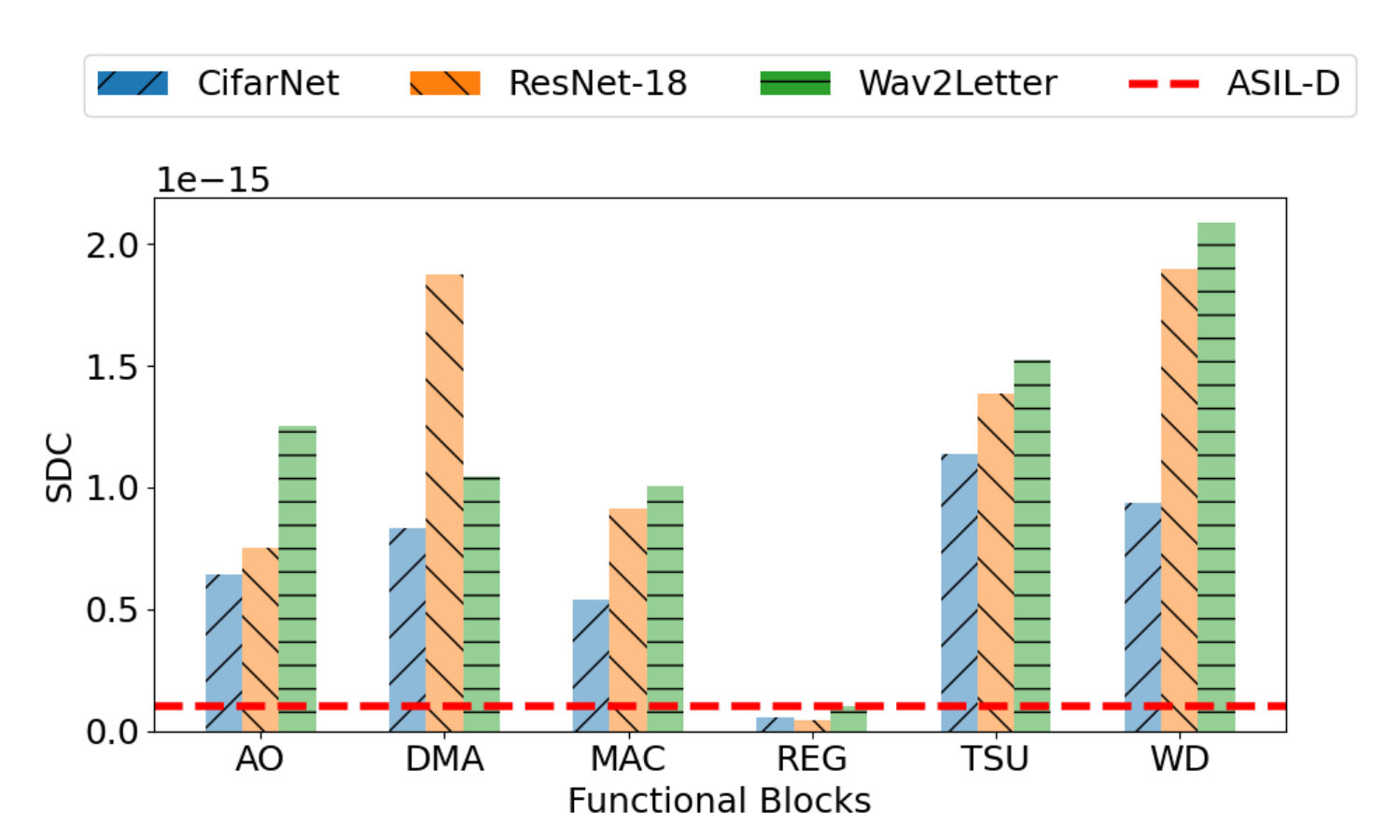}
  \caption{Block-wise SDC contribution for different applications running on Arm Ethos-U55~\cite{u55} with MAC-32 configuration.}
  \label{fig:sens_app}
\end{figure}

To verify our intuition, we carry out RTL fault injection on Ethos-U55 with three different sets of applications. \Fig{fig:sens_app} shows that blocks like DMA might be less resilient than AO while running ResNet-18, but that might not be the case when CifarNet is running on U55. Therefore, if a protection scheme is devised with just few applications in mind, it might fall short of the required resiliency levels for other applications.

\subsubsection{Sensitivity to Technology Node}
\label{sens:node}

A chosen technology node can dictate the soft-error reliability of a single FF~\cite{narasimham2018scaling} and hence that of functional blocks comprising the FFs. ~\cite{seifert2015soft} et.al show how FFs Soft Error Rates (SER) have reduced drastically with advanced technology nodes, which makes them less susceptible to soft errors. However, it has been observed that the soft error rates of FFs resulting from faults in combinational logic elements have increased as technology nodes advance. Consequently, if an NPU design necessitates fabrication with a newer technology node, a soft-error resilient scheme tailored to the behavior of functional blocks in an older technology node may not be optimal. 

To further illustrate these findings, Fig.~\ref{fig:sens_node} depicts the variation in the soft-error reliability of Ethos U55~\cite{u55} functional blocks in 16nm and 7nm Bulk FinFET technologies. We calculate the SDCs for each functional block as described in ~\Sect{sec:comb} and use the Soft Error Rate (SER) FIT values for the two technologies as mentioned in prior work~\cite{cao2019alpha}. 

\begin{figure}[t]
  \centering
  \includegraphics[width=\columnwidth]{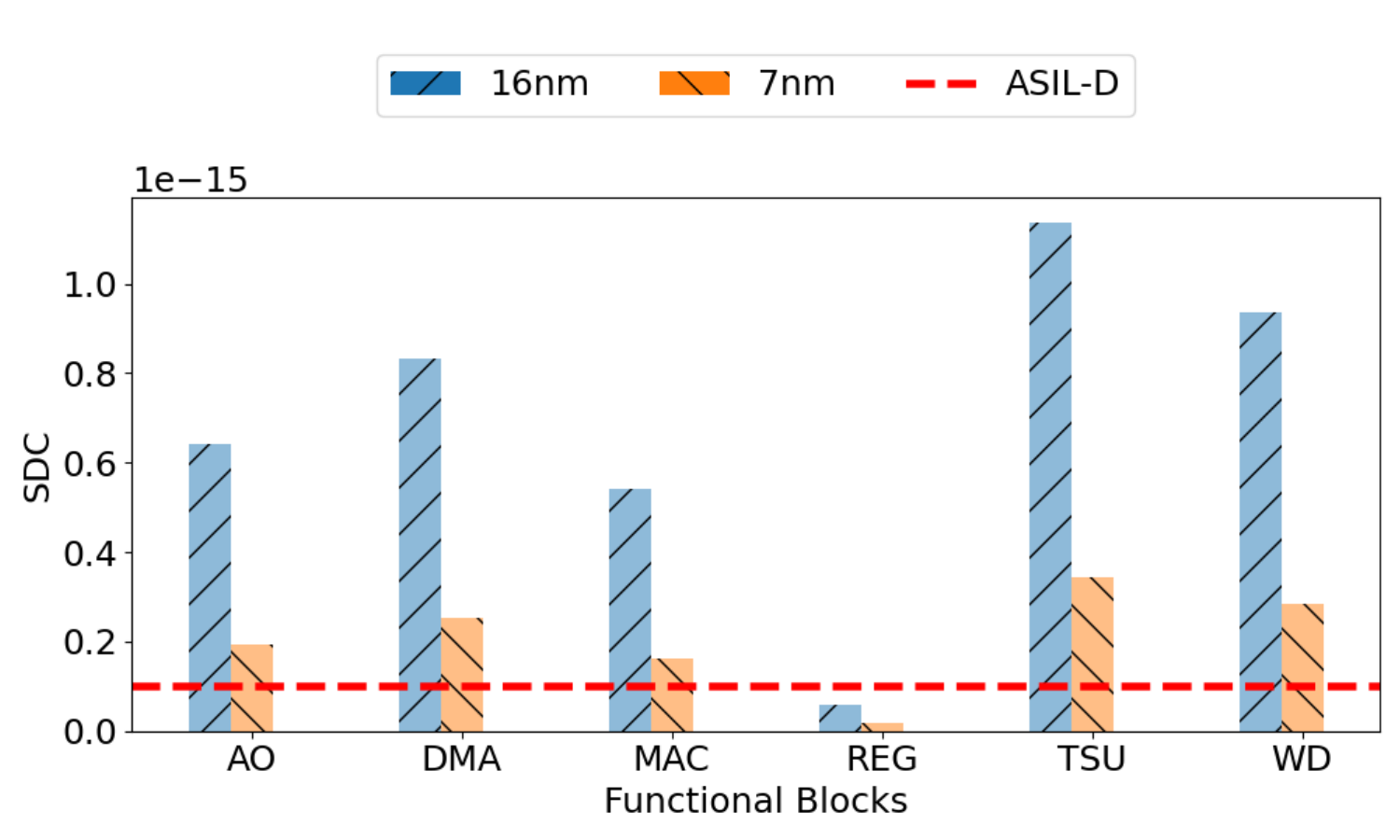}
  \caption{Variation in the reliability of Arms Ethos U55 \mbox{\cite{u55}}, for TSMC 16 nm and 7 nm technology nodes for MAC-32 configuration running ResNet-18.}
  \label{fig:sens_node}
\end{figure}

The SDC rates of the functional blocks in 7nm is on average 3.3$\times$ less than 16nm. This resiliency behavior over technology nodes is down to factors such as the sensitive area of a storage cell, Critical Charge ($Q_{crit}$) and Collected Charge ($Q_{coll}$). For the 7nm FinFET node, the amount of charge collected, i.e. $Q_{coll_{7nm}}$ is less than $Q_{coll_{16nm}}$ which results in higher SER FIT rate for the FF in 16nm technology node~\cite{cao2019alpha}.

\subsubsection{Combinational Logic Faults.}
\label{sens:logic}

Neglecting faults in combinational logic elements leads to an overestimation of reliability. Prior works mostly ignore combinational logic faults because soft errors in FF and memory are present for a longer time whereas a Single Event Transient (SET) generated at the output of a combinational element affects the system only if it gets latched by a FF. Previously, multiple levels of masking~\cite{shivakumar2002modeling} has rendered such a case unlikely. However, with technology and voltage downscaling and increasing clock frequency, the total contribution of SETs to Soft Error Rates (SERs) has increased beyond negligible\mbox{\cite{mahatme2014impact}}. 

We show in ~\Fig{fig:sens_log} the difference in reported SDC contribution of each functional block in Ethos-U55, for the cases when logic faults are and are not considered (See~\Sect{sec:comb} for logic fault SDC contribution methodology).
Clearly, the reliability of all functional blocks is lower when combinational faults are considered, showing that studying logic faults is warranted.
The sensitivity of functional block SDC rate to logic faults consideration adds another variable to the search of an optimal soft-error resilient scheme. 
\begin{figure}[t]
	\centering
	\includegraphics[width=\columnwidth]{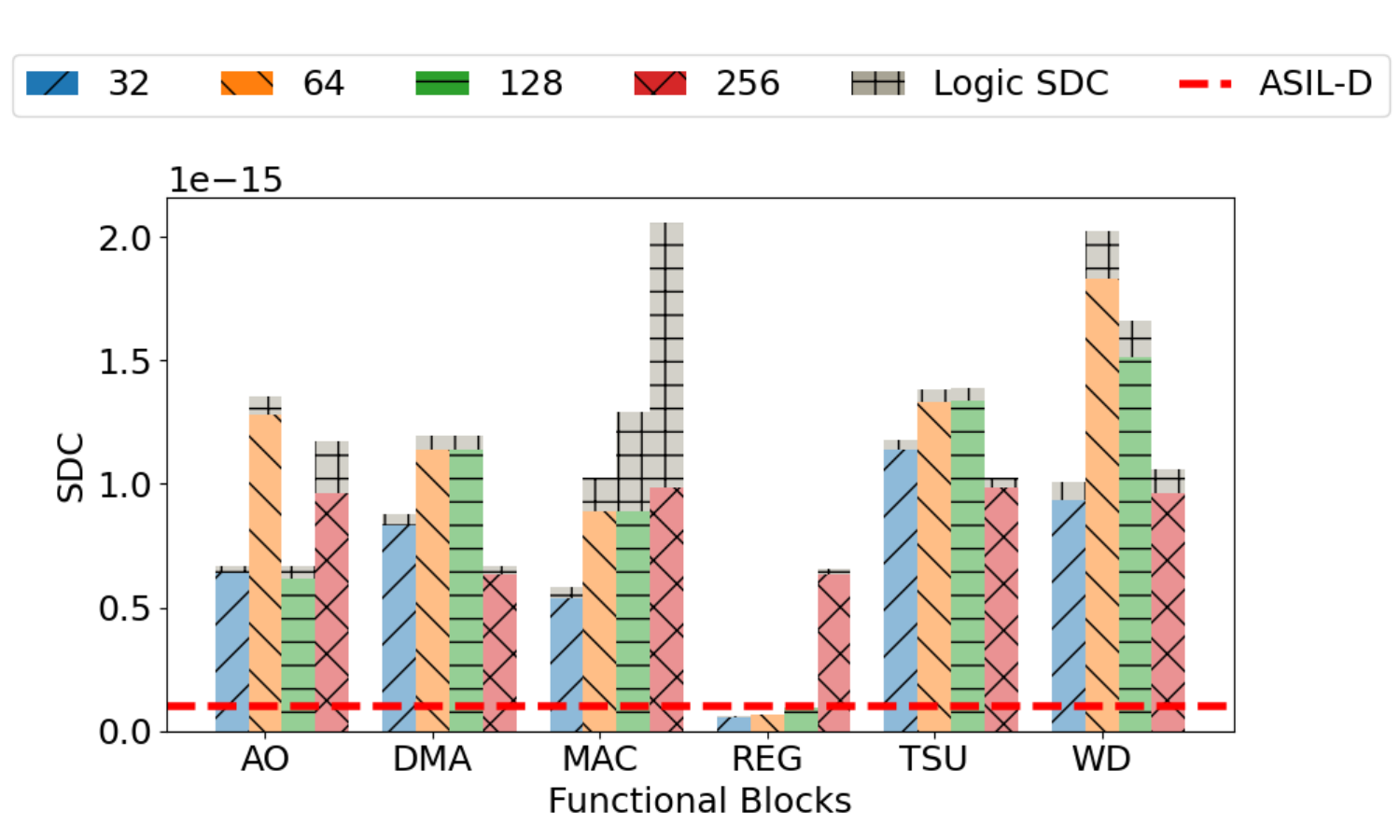}
	\caption{Block-wise SDC contribution of Arm Ethos-U55\mbox{\cite{u55}}, while considering and not considering logic faults for all the four MAC configurations in TSMC 16 nm technology node running ResNet-18.}
\label{fig:sens_log}
\end{figure}
\subsection{Area vs SDC Tradeoff Analysis for Ethos-U55}
\label{sec:sdc_area}
\begin{figure}[t]
	\centering
	\includegraphics[width=\columnwidth]{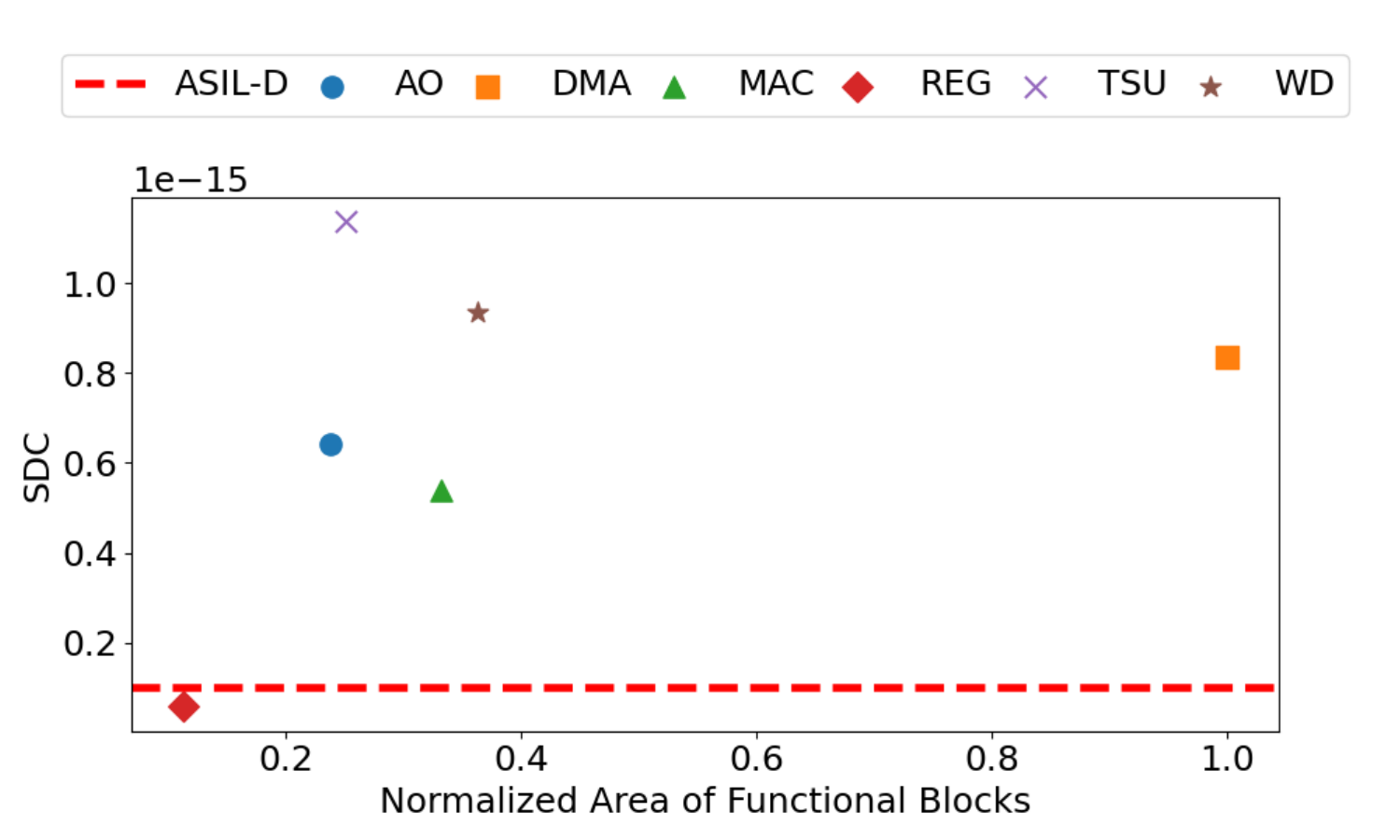}
	\caption{Block-wise SDC contribution of Arm Ethos-U55 for MAC-32 configuration while running ResNet-18.}
	\label{fig:func_block}
\end{figure}
Our experiments have shown that the resiliency of the NPU is a function of numerous factors interacting in a non-trivial manner. With Ethos-U55 being utilised in safety-critical applications, it becomes crucial to understand how a resilient version of Ethos-U55 can be designed with existing soft-error mitigation and detection strategies to meet safety standards. 

\Fig{fig:func_block} portrays the spectrum of SDC rates and corresponding area footprint of NPUs functional blocks for MAC-32 configuration running a CifarNet~\cite{hosang2015taking}.
Functional blocks such as Clock and Power Module (CPM) produces no SDCs (hence are absent from the plot) as the block generates a clock for the IP, not affecting any computation (they lead to crashes).
Traversal Unit (TSU) is the most vulnerable block due to its function: TSU is part of CC, which manages the order of execution and traverses the inputs correctly for the output-stationary data-flow.  

Critically, the number of FF (or area) of a functional block is in no way an indication of their inherent soft-error resiliency behavior.
Intuitively, resiliency of a block is owing to multiple factors such as the design, dataflow, utilization of the block, workload, etc.
Prior work~\cite{papadimitriou2023avgi} has showed this trend for traditional CPUs using the variation in Architectural Vulnerability Factor (AVF)~\cite{leveugle2009statistical} of functional components such as L1 Cache, Physical Register File, and Reorder Buffer. 

For instance, AO and TSU, despite having almost the same area, differ drastically in their inherent soft-error resiliency.This is because AO is responsible for applying non-linear activations to the output computed by MAC unit.
Therefore, the effect of a bit-flip in AO is likely to be masked by either the non-linear activation function or the approximate nature of neural networks~\cite{chan2021understanding,hoang2020ft}.  
Similarly, DMA has 2$\times$ the area of WD but the two share a similar soft-error resilience. This is because of DMA's sporadic use in U55, as the off-chip data movement is very limited due to high reuse.

The absence of a positive correlation between the block area and its resiliency provides an opportunity to understand which blocks to make more resilient to meet system resiliency requirements under area constraints. 


\section{Understanding Ethos-U55's Resiliency Under Existing Soft-Error Mitigation Techniques}
\label{sec:ext_se}

We first introduce the SDC rate per inference of the NPU ($SDC_{NPU}$) formulation (Sec~\ref{sec:cont:form}) using an example hardware with just three fault sites. We then show how $SDC_{NPU}$ can be formulated as a function of SDC contribution of various functional blocks (Sec~\ref{sec:fblock}). We then introduce how $SDC_{NPU}$ can be estimated accurately and feasibly (Sec~\ref{sec:sdc_est}). And lastly, we explain how $SDC_{NPU}$ can be calculated when combinational faults are considered (Sec~\ref{sec:comb}).

\subsection{$SDC_{NPU}$ Formulation}
\label{sec:cont:form}
With the varying level of functional block level resiliency, the search space for finding an optimal soft-error mitigation scheme is a vast one, which requires solving the following constrained optimization problem,

\begin{mini}|l|
	{}{SDC_{NPU}}{}{}
	\addConstraint{area}{\leq a_{budget}}{}
	\label{eq:opt}
\end{mini}

\noindent where $SDC_{NPU}$ is the SDC rate per inference of the NPU, and can be calculated by particle beam experiments~\cite{li2023seu}, RTL fault injection~\cite{tyagi2022thales} or modelling the hardware error behavior~\cite{mahmoud2020pytorchfi}.

\begin{figure}[t]
	\centering
	\includegraphics[width=\columnwidth]{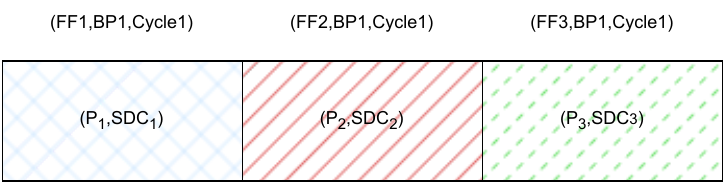}
	\caption{An illustration of hardware fault-site (i.e., a bit position of a chosen FF at a particular cycle). Each fault site is characterized by its probability to experience a bit flip ($P_{i}$) and the SDC rate of the fault site ($SDC_{i}$).}
\label{fig:3fs}
\end{figure}

Our idea is illustrated in Fig~\ref{fig:3fs}, where each box represents a hardware FF or fault-sites (we do not require FF to be in close proximity. In a real system, multiple bit-flips can occur in FF which may or may not belong to one functional block). Each fault site is represented by two parameters:
\begin{itemize}
\item $P_{i}$: is the probability that a bit-flip occurs at fault-site $i$ at any instant of time and $P^{'}_{i} = (1 - P_{i})$.
\item $SDC_{i}$: is the SDC rate of the system, when a bit-flip occurs at fault-site $i$. 
\end{itemize}

The probability $P_{i}$ is dictated by the raw FIT rate $FIT_i$ of the FF $i$. Raw FIT rate gives the total number of failures, i.e., bit-flips, expected in the FF in 1 Billion hours of operation. Hence, the probability that a failure can occur at any instant of time (at a cycle), can be written as:

\begin{align}
	P_{i} = \frac{FIT_{i}}{{2^{20}}\times 8 \times 10^9 \times 3600 \times freq}
	\label{eq:prob}
\end{align}

\noindent where FIT rate of a FF is given as \textit{FIT/MB} and $freq$ is the frequency of operation of the NPU.\\

The central question is, what is the SDC rate of the NPU if we know the probabilities and SDC rate of \textit{each} fault site?
The crux is to consider all possible events when a particle strike happens at the hardware and model how SDC of each fault-site contributes to $SDC_{NPU}$. Consider Fig~\ref{fig:3fs} as a representative hardware with three fault sites. If a particle strike happens on this hardware, the resulting behavior of the hardware can be categorized as either of the following three categories:
\begin{itemize}
	\item \textbf{Single Bit-Flip}: In this event, a bit-flip occurs only at one fault site i.e., either at fault-site $FS_{1}$, $FS_{2}$ or $FS_{3}$.
	\item \textbf{Multiple Bit-Flips}: In this case, more than one underlying flip-flops can sustain bit-flips. Which means affected fault-sites could be $FS_{1}FS_{2}$, $FS_{1}FS_{3}$, $FS_{2}FS_{3}$ or $FS_{1}FS_{2}FS_{3}$.
	\item \textbf{No Bit-Flip}: And finally, there could be a case where no bit-flip occurs in the hardware. 
 \end{itemize}
 
We assume that the occurrence or non-occurrence of any event does not affect the probability of other events happening. This is a fair assumption to make as we can see from Equ~\ref{eq:prob}, the probability of a fault occurring in a FF is dependent on the raw FIT rate, which is an intrinsic property of the FF. In that case, with basic probability theory, we can calculate the probability of all possible 8 events in the case of given three fault sites.

\begin{itemize}
\item {\textit{(Case 1)} Fault occurs at fault-site 1:} $P_{1}P^{'}_{2}P^{'}_{3}$
\item {\textit{(Case 2)} Fault occurs at fault-site 2:} $P^{'}_{1}P_{2}P^{'}_{3}$
\item {\textit{(Case 3)} Fault occurs at fault-site 3:} $P^{'}_{1}P^{'}_{2}P_{3}$
\item {\textit{(Case 4)} Fault occurs at fault-site 1 and 2:} $P_{1}P_{2}P^{'}_{3}$
\item {\textit{(Case 5)} Fault occurs at fault-site 1 and 3:} $P_{1}P^{'}_{2}P_{3}$
\item {\textit{(Case 6)} Fault occurs at fault-site 2 and 3:} $P^{'}_{1}P_{2}P_{3}$
\item {\textit{(Case 7)} Fault occurs at fault-site 1,2 and 3:} $P_{1}P_{2}P_{3}$
\item {\textit{(Case 8)} No fault occurs:}$P^{'}_{1}P^{'}_{2}P^{'}_{3}\\$
\end{itemize}

Having worked out the probabilities of all possible events, $SDC_{NPU}$ can be written as a sum of the SDC contribution from each possible event. In other words,

\begin{equation}
\begin{split}
SDC_{NPU} =  &P_{1}P^{'}_{2}P^{'}_{3}SDC_{1}+P^{'}_{1}P_{2}P^{'}_{3}SDC_{2}+P^{'}_{1}P^{'}_{2}P_{3}SDC_{3}+\\
&P_{1}P_{2}P^{'}_{3}SDC_{12}+P_{1}P^{'}_{2}P_{3}SDC_{13}+P^{'}_{1}P_{2}P_{3}SDC_{23}+\\
&P_{1}P_{2}P_{3}SDC_{123}+P^{'}_{1}P^{'}_{2}P^{'}_{3}SDC_{none}
\label{eq:sdcnpu}
\end{split}
\end{equation}

\noindent where $SDC_{i}$ is the SDC of the system when only the $i^{th}$ event occurs.
If we look at Equ~\ref{eq:sdcnpu}, we can simplify it further.
Firstly, $SDC_{none} = 0$ as there are no SDCs to be observed when no fault occurs.
Secondly, since the order of $P_{i} \approx 10^{-18}$, we can approximate $P^{'}_{i} \approx 1$.
Lastly, the probability of multiple-bit flips taking place is a product of two or more probabilities, which ranges from $10^{-36}$ to $10^{-54}$. With such low probabilities, it can be safely assumed that the likelihood of such an event occurring is negligible, hence simplifying Equ~\ref{eq:sdcnpu} as:

\begin{align}
		SDC_{NPU} \approx P_{1}SDC_{1} + P_{2}SDC_{2} +	P_{3}SDC_{3}
		\label{eq:sdcnpu_simpl}
\end{align}
\subsection{$SDC_{NPU}$ formulated as functional block SDC}
\label{sec:fblock}

If we look at Equ~\ref{eq:sdcnpu}, we can simplify it further. For instance, firstly, $SDC_{none} = 0$ as there are no SDCs to be observed when no fault occurs. Secondly, since the order of $P_{i} \approx 10^{-18}$, we can approximate $P^{'}_{i} \approx 1$. And lastly, for \textit{(Case 4)} to \textit{(Case 7)}, the probabilities of the event taking place is a product of two or more probabilities, the order of which ranges from $10^{-36}$ to $10^{-54}$. With such low probabilities, it can be safely assumed that the likelihood of such an event occurring is negligible, hence simplifying Equ~\ref{eq:sdcnpu} as:

\begin{align}
		SDC_{NPU} \approx P_{1}SDC_{1} + P_{2}SDC_{2} +	P_{3}SDC_{3}
		\label{eq:sdcnpu_simpl}
\end{align}

Extending our three fault site example to a hardware with $N$ fault sites, we can generalize Equ~\ref{eq:sdcnpu_simpl} as:

\begin{align}
	SDC_{NPU} \approx \sum_{i=1}^{N}P_{i}SDC_{i}
	\label{eq:sdcnpu_gen}
\end{align}

Intuitively, Equ~\ref{eq:sdcnpu_gen} is just the summation of the product probability of a fault occurring at fault-site $i$ and the SDC rate of the fault-site.  

As we are specifically interested in quantifying the reliability of an NPU such as in Fig~\ref{fig:npu}, we can re-write

\begin{align}
	N = N_{CC} + N_{DMA} + N_{AU} + N_{WU} + N_{MAC} + N_{OU}
	\label{eq:fs}
\end{align}

\noindent where $N_{K}$ is the total number of fault-sites in functional block $K$. We can make use of Equ~\ref{eq:fs} to re-write $SDC_{NPU}$ as 

\begin{equation}
\begin{split}
	SDC_{NPU} \approx &\sum_{i=1}^{N_{CC}}P_{CC}SDC_{i}+\sum_{j=1}^{N_{DMA}}P_{DMA}SDC_{j}+\\
							&\sum_{k=1}^{N_{AU}}P_{AU}SDC_{k}+\sum_{l=1}^{N_{WU}}P_{WU}SDC_{l}+\\
							&\sum_{m=1}^{N_{MAC}}P_{MAC}SDC_{m}+\sum_{n=1}^{N_{OU}}P_{OU}SDC_{n}
	\label{eq:sdcnpu_block}
\end{split}
\end{equation}

Equ~\ref{eq:sdcnpu_block} can be interpreted as writing $SDC_{NPU}$ as the contribution of SDC from each functional block. A key observation to make is that for any of the functional blocks, we assume that the probability of a fault occurring in a fault-site within the block is uniform, hence the omission of $P_{i}, P_{j}....P_{n}$ from Equ~\ref{eq:sdcnpu_block}. This is a fair assumption to make, as for our purposes we assume that a chosen protection/detection scheme is applied to the entirety of a functional block.

\subsection{Estimating $SDC_{NPU}$}
\label{sec:sdc_est}

Equ~\ref{eq:sdcnpu_gen} calculates $SDC_{NPU}$ precisely. However, it is impractical to calculate that equation because of the large number of fault sites $N$. Calculating $SDC_{i}$ requires running RTL simulations for each fault site over the entire test set (MobileNet \cite{howard2017mobilenets} has more than 1 billion fault sites).

We use \textit{THALES}~\cite{tyagi2022thales}, a reliability estimation tool validated against RTL fault injection to estimate the $SDC_{NPU}$ as per our formulation in Sec~\ref{sec:cont:form} for all the possible configurations of Ethos-U55. We observe that SDC calculation can be formulated as integrating a discrete function over a finite domain:

\begin{equation}
	\begin{split}
	&SDC_{NPU} = \sum_{j=1}^{N}P_{j}SDC_{j},~~~j \in \mathbb{Z} : j \in [1, N]
	\label{ra:sdcnpu_mc}
	\end{split}
\end{equation}

In Equ~\ref{ra:sdcnpu_mc}, integrand $f(\cdot)$ does not have an analytical form that can be calculated in practice. In such a case, we propose to solve the integration numerically using Monte Carlo integration~\cite{press2007numerical}. Formally, $SDC_{NPU}$ can be estimated by drawing $K$ independent samples using a Probability Density Function(PDF) and calculate:

\begin{align}
	\overline{SDC_{NPU}} =  \frac{1}{K}\sum_{j=1}^{K}\frac{f(X_j)}{PDF(X_j)},~~~\sum_{j=1}^{N}PDF(X_j) = 1
\end{align}

\noindent where $\overline{SDC_{NPU}}$ is the Monte Carlo Estimator of $SDC_{NPU}$. For our purposes, we chose $PDF = \frac{1}{N}$, where we sample the fault-space uniformly. With the PDF selected for our Monte Carlo Estimator of $SDC_{NPU}$, we can estimate Equ~\ref{eq:sdcnpu_gen} as

\begin{align}
	\overline{SDC_{NPU}} = \frac{N}{K}\sum_{i=1}^{K}P_{i}SDC{i}
	\label{sdc:gen_samp}
\end{align}

\noindent where $K$ is the total number of independent samples drawn from all the fault sites. As we are interested in the resiliency characteristics of functional blocks, we can re-write the Equ~\ref{sdc:gen_samp} using Equ~\ref{eq:fs} as follows:
\begin{equation}
\begin{split}
	\overline{SDC_{NPU}} = &P_{CC}\times \frac{N_{CC}}{K_{CC}}\sum_{i=1}^{K_{CC}}SDC_{i}+P_{DMA}\times \frac{N_{DMA}}{K_{DMA}}\sum_{j=1}^{K_{DMA}}SDC_{j}+\\
	&P_{AU}\times \frac{N_{AU}}{K_{AU}}\sum_{k=1}^{K_{AU}}SDC_{k}+P_{WU}\times \frac{N_{WU}}{K_{WU}}\sum_{l=1}^{K_{WU}}SDC_{l}+\\
	&P_{MAC}\times \frac{N_{MAC}}{K_{MAC}}\sum_{m=1}^{K_{MAC}}SDC_{m}+P_{OU}\times \frac{N_{OU}}{K_{OU}}\sum_{n=1}^{K_{OU}}SDC_{n}
	\label{eq:sdcnpu_block_samp}
\end{split}
\end{equation}

\noindent with $K_{functional-block}$ being the total number of independent samples drawn from the \textit{functional block}. 

Equ~\ref{eq:sdcnpu_block_samp} clearly articulate the respective contributions of each functional block to the overall ${SDC_{NPU}}$. Consequently, it serves as a valuable tool for evaluating the potential impact of employing specific soft-error mitigation strategies within individual functional blocks or in combination.

\subsection{Estimating $SDC_{NPU}$ With Logic Faults}
\label{sec:comb}
The ${SDC_{NPU}}$ modeling so far ignores logic faults.
When logic faults are taken into consideration, the probability that a bit-flip occurs in a FF is higher than that when logic errors are ignored.
We model this behavior as an increase in the FIT rate of a FF.
Specifically, we can modify the Equ~\ref{eq:prob} to:


\begin{align}
	P_{i} = \frac{FIT_{i}^{'}}{2^{20}\times 8 \times 10^9 \times 3600 \times freq}\\
	FIT_{i}^{'} = FIT_{i} + \alpha
	\label{eq:prob_logic_alpha} 
\end{align}

\noindent where $\alpha$ is the factor by which FIT rate of a FF increases, and according to Seifert et al.~\cite{seifert2010radiation}, is calculated as
\begin{align}
	\alpha = SER_{Comb} \times \frac{10^9}{T}
\end{align}
\begin{align}
	SER_{Comb} = flux \times CrossSectionArea
\end{align}

\noindent where $T$ is the number of hours of operation, $SER_{Comb}$ $(error/hr)$ is the SER from the logic, $flux$ describes the amount of particles that are bombarded per unit area of silicon per unit time ($particles/cm^2/hr$), and \textit{Cross Section Area} ($cm^2$) is the logic gate area that is sensitive to charged particles.

An upper bound of $SER_{Comb}$ can be calcualted by using the methodology described by Gill et al.~\cite{gill2009comparison}, with the values described in ~\Tbl{tbl:comb_logic}.
The formulation describes $SER_{Comb}$ as a percentage of nominal latch $SER$ and is calculated as:

\begin{equation}
	\begin{aligned}
		& \frac{\mathrm{SER}_{\text {comb }}}{\mathrm{SER}_{\text {Latch }}}\% \approx \mathrm{LD}_{\text {comb }} * freq * \\
		& *\left(FOM *\left\{\begin{array}{l}
			\frac{\left(\text { Fanin }^{<\mathrm{d}>+1}-1\right)}{\left(\text { Fanin-1}\right)} ; \text { Fanin }>1 \\
			<\mathrm{d}>\text {; Fanin }=1
		\end{array}\right)\right. \\
		&
		\label{eq:ser_comb}
	\end{aligned}
\end{equation}

\noindent where Figure of Merit (FOM) is a technology and frequency-dependent parameter. Since we are calculating an upper bound on $SER_{Comb}$, $LD_{Comb} = 1$ (all logic faults reach a FF to get captured), frequency = 1GHz, and $SER_{Latch} = \frac{1}{2}SER_{FF}$~\cite{gill2009comparison}, where $SER_{FF}$ is calculated for each block. (See ~\Sect{sec:se_char}).

Fanin is the average number of inputs to logic gates in the circuit. The higher the fan-in the larger the number of susceptible logic gates at a fixed depth <d> feeding into one equivalent latch. Since we only consider alpha-particles in our study, we use d = 3.5 as mentioned by Gill et al.~\cite{gill2009comparison}. We calculate the average Fanin for our four MAC configurations by using the netlist obtained after synthesis and using the all-fanin command available in Synopsys Design Compiler.

\begin{table}[t]
	\centering
	\caption{Conditions for estimating logic faults at different technology nodes. [Flux($particles/cm^2/hr$) = 0.001].} 
	\renewcommand*{\arraystretch}{1}
	\renewcommand*{\tabcolsep}{4pt}
	\resizebox{\columnwidth}{!}
	{
		\begin{tabular}{cccccc}
			\toprule[0.15em]
			\textbf{\specialcell{Technology\\ Node}} & \textbf{\specialcell{Voltage\\ ($V$)}}  & \textbf{\specialcell{FOM (\%)}~\cite{xiong2022single}} & \textbf{\specialcell{Cross Section\\ Area ($cm^2$)~\cite{harrington2019empirical}}}  & \textbf{\specialcell{Critical\\ Charge ($fC$)}} & \textbf{\specialcell{FF FIT\\ Rate (FIT/MB)}}\\
			\midrule[0.05em]
			16 nm       & 0.75   & 0.5  & $3 \times 10^-11$ & 0.9477 & 50\\
			7 nm       & 0.7  & 0.05  &  $0.306 \times 10^-11$ & 0.8059 & 10\\ 
			\bottomrule[0.15em]
		\end{tabular}
	}
	\label{tbl:comb_logic}
\end{table}


\section{Ethos-U55 Resiliency Improvements Under DMR and Flop Hardening}
\label{sec:u55_dmr}
\begin{figure*}[t]
	\centering
	\subfloat[\small{DMR Without Logic Faults}]
	{
		\includegraphics[trim=0 0 0 0, clip, width=0.6\columnwidth]{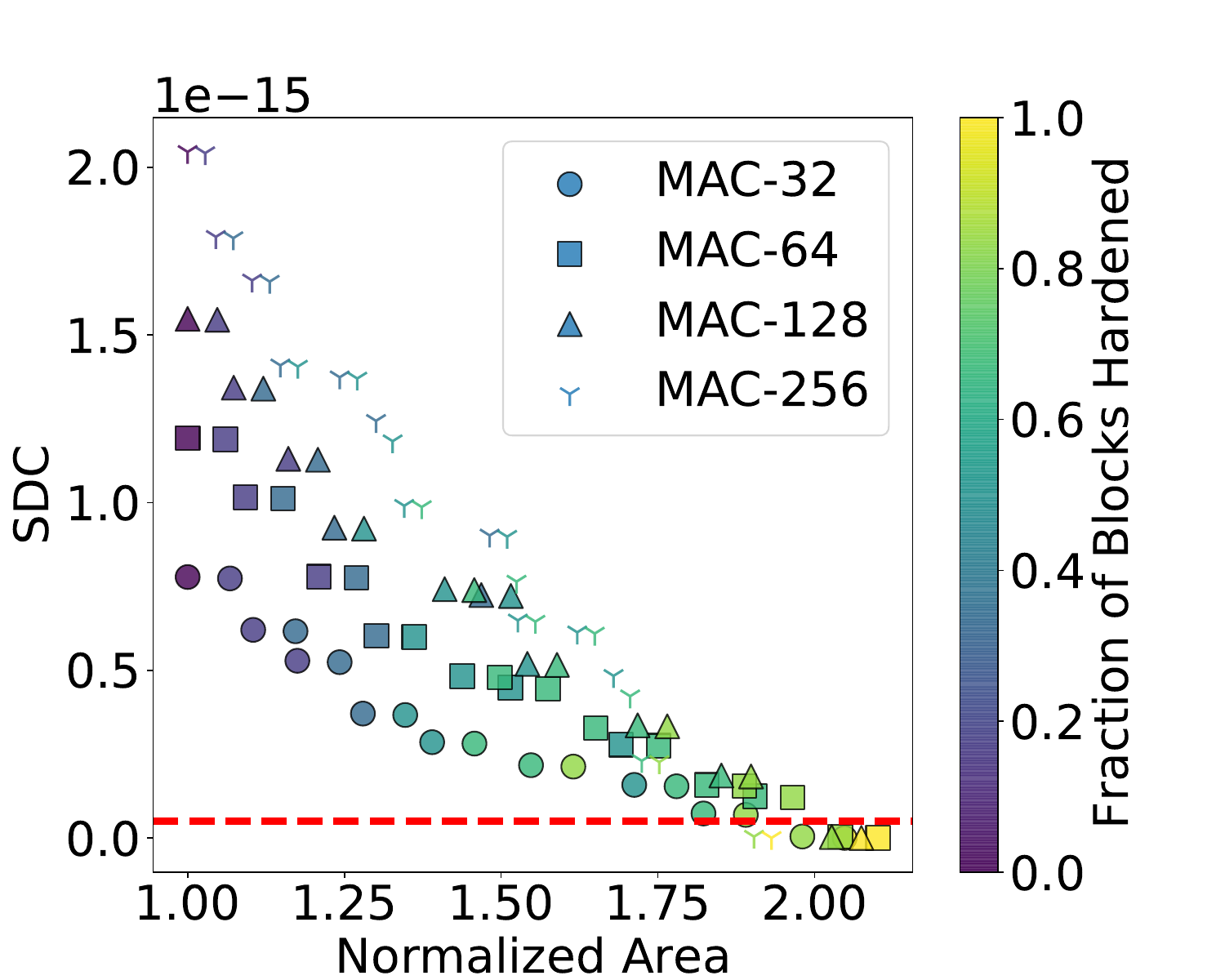}
		\label{fig:u55_dmr}
	}
	\subfloat[\small{DMR With Logic Faults}]
	{
		\includegraphics[trim=0 0 0 0, clip, width=0.6\columnwidth]{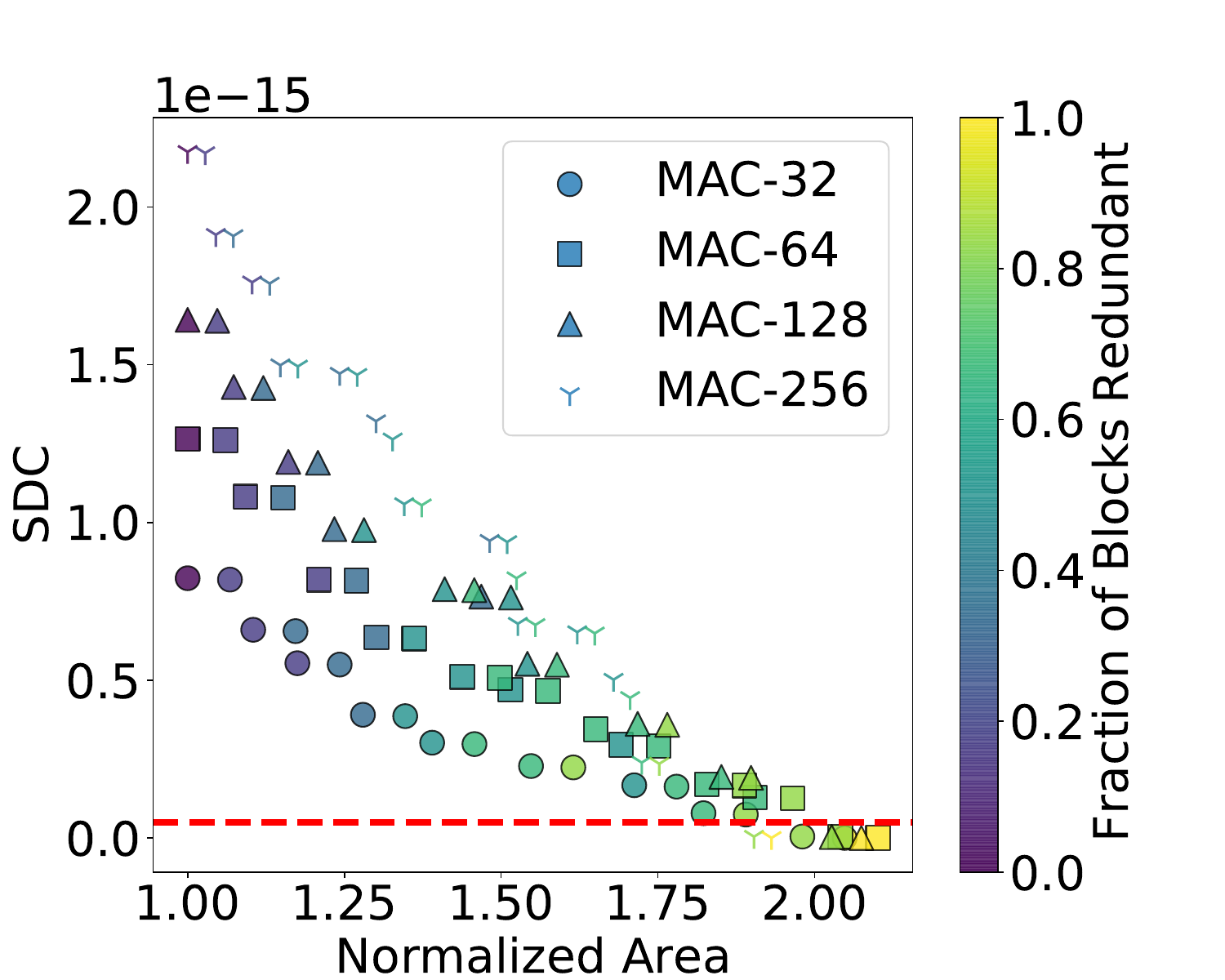}
		\label{fig:u55_dmr_logic}
	}
	\subfloat[\small{Flop Hardening Without Logic Faults}]
	{
		\includegraphics[trim=0 0 0 0, clip, width=0.6\columnwidth]{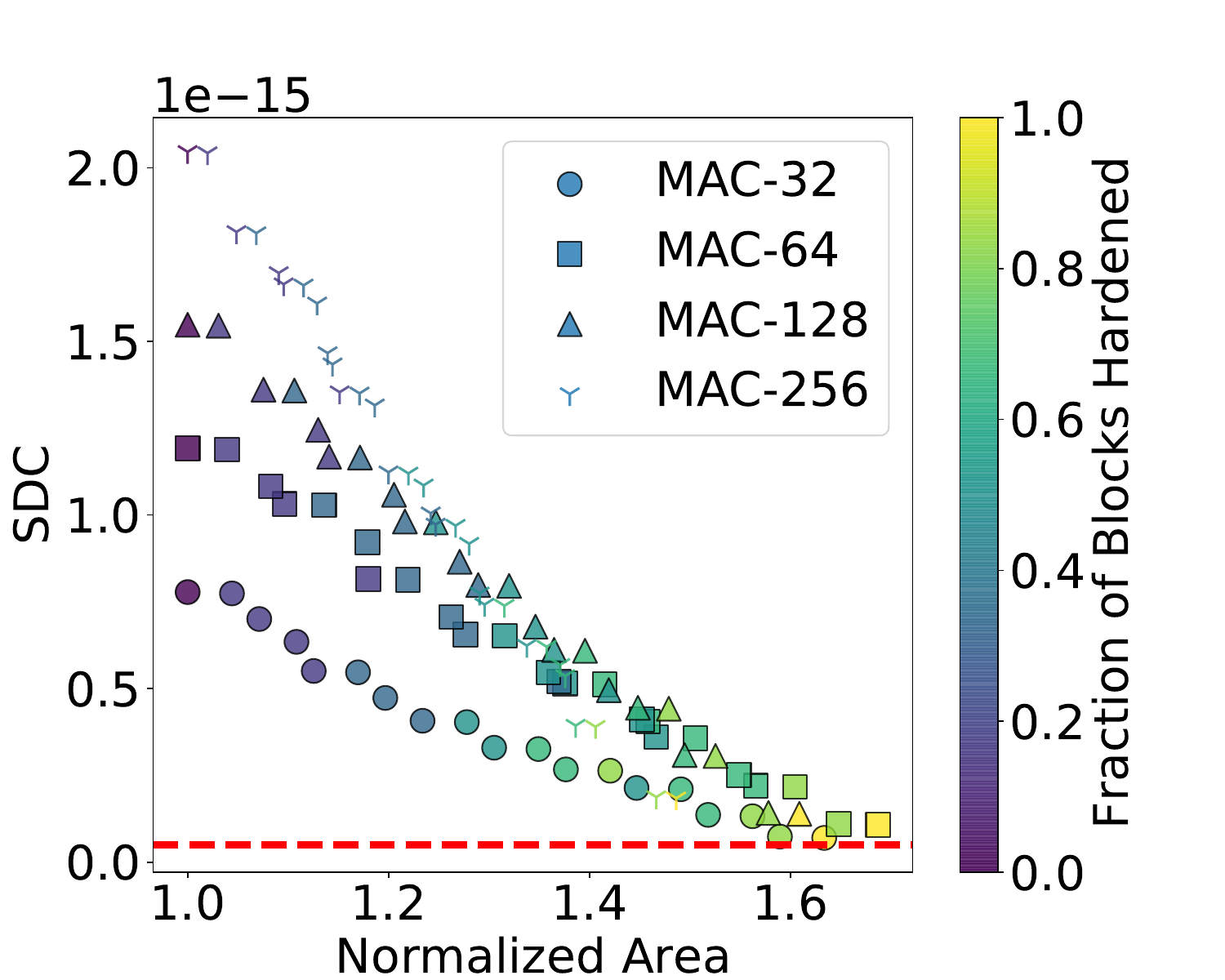}
		\label{fig:u55_hard}
	}
	\hspace{1pt}
	\subfloat[\small{Flop Hardening With Logic Faults}]
	{
		\includegraphics[trim=0 0 0 0, clip, width=0.6\columnwidth]{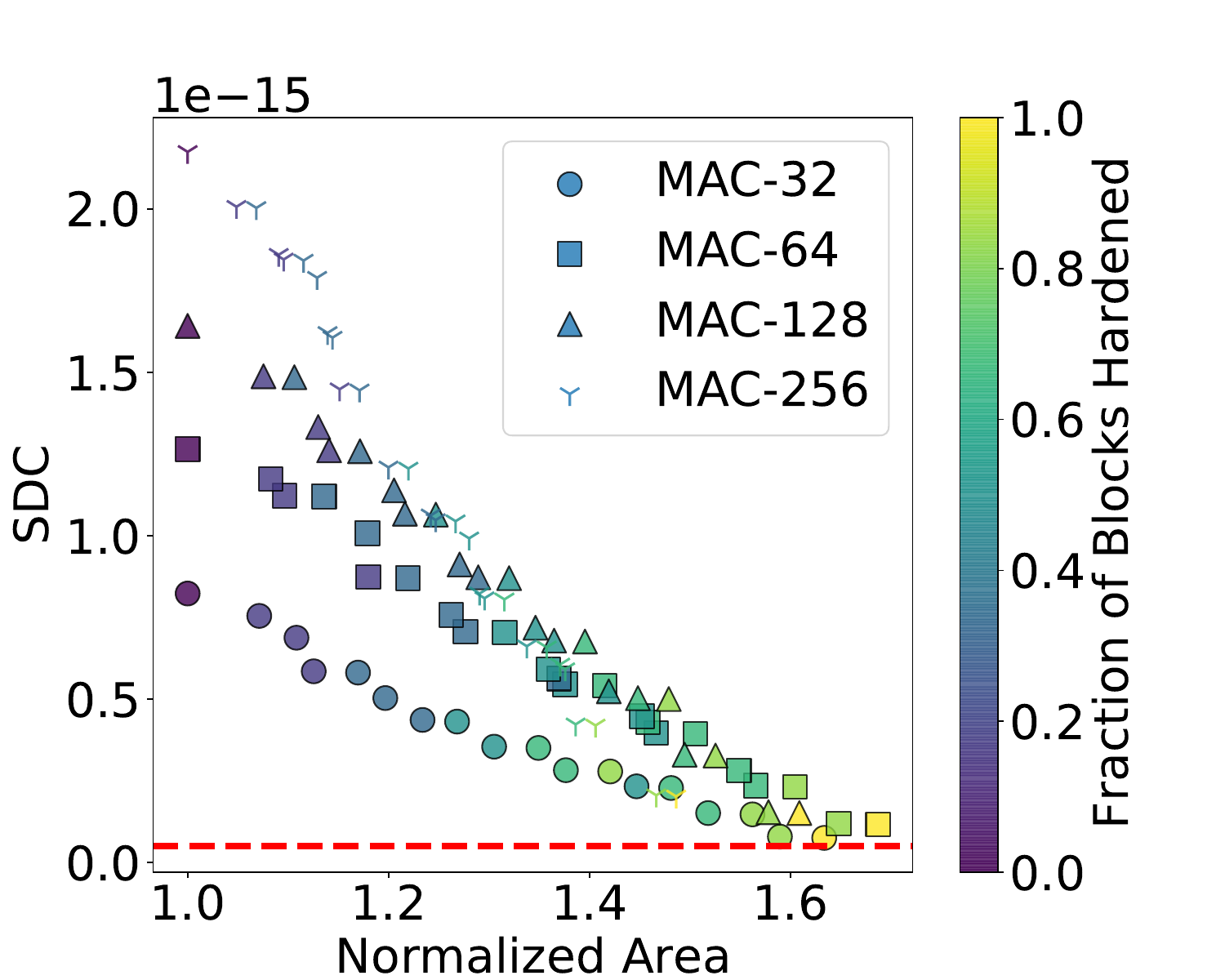}
		\label{fig:u55_hard_logic}
	}
	\subfloat[\small{DMR, Flop Hardening + SET }]
	{
		\includegraphics[trim=0 0 0 0, clip, width=0.6\columnwidth]{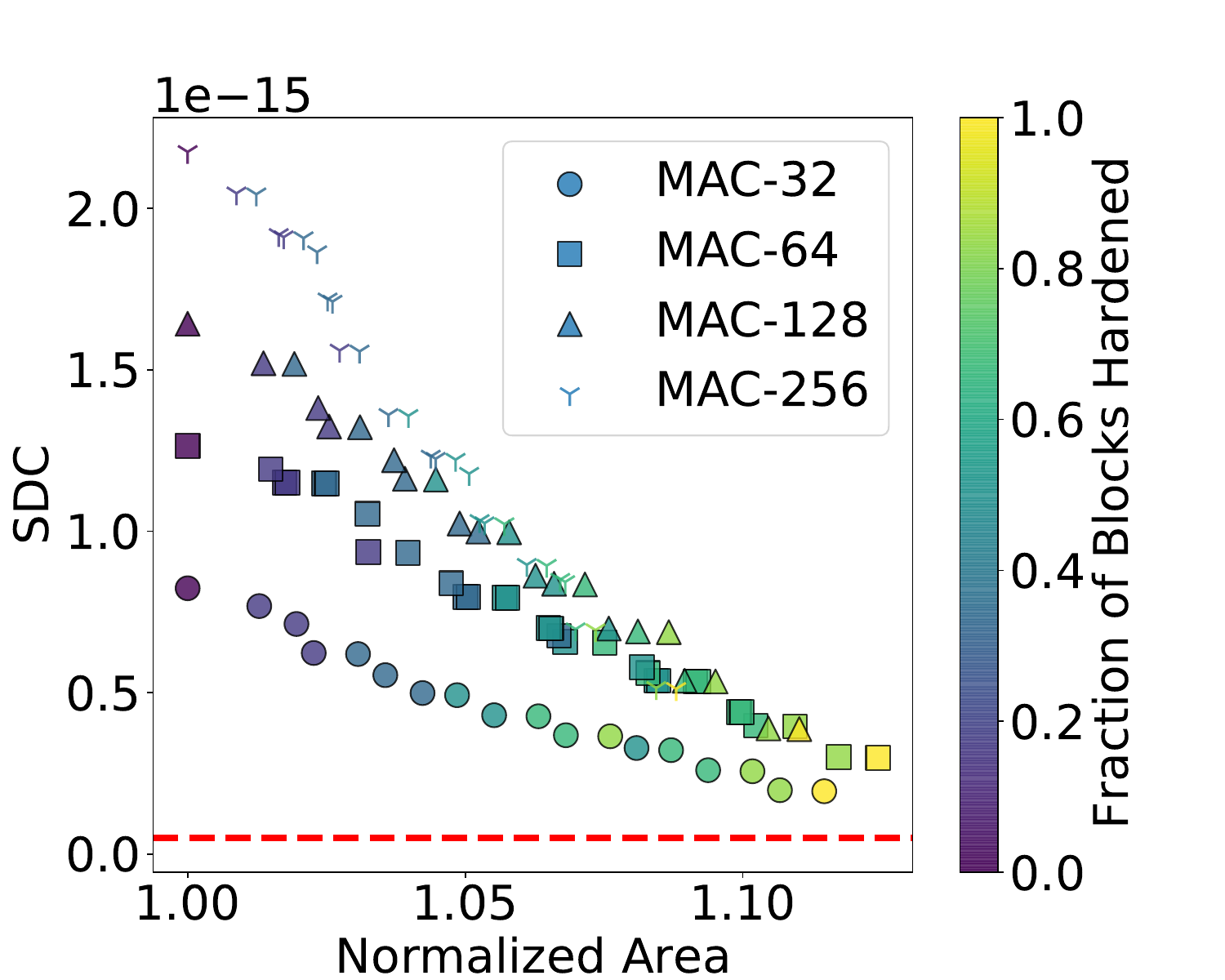}
		\label{fig:u55_hard_set}
	}
	\subfloat[\small{Flop Hardening + SET W/ Logic Faults}]
	{
		\includegraphics[trim=0 0 0 0, clip, width=0.6\columnwidth]{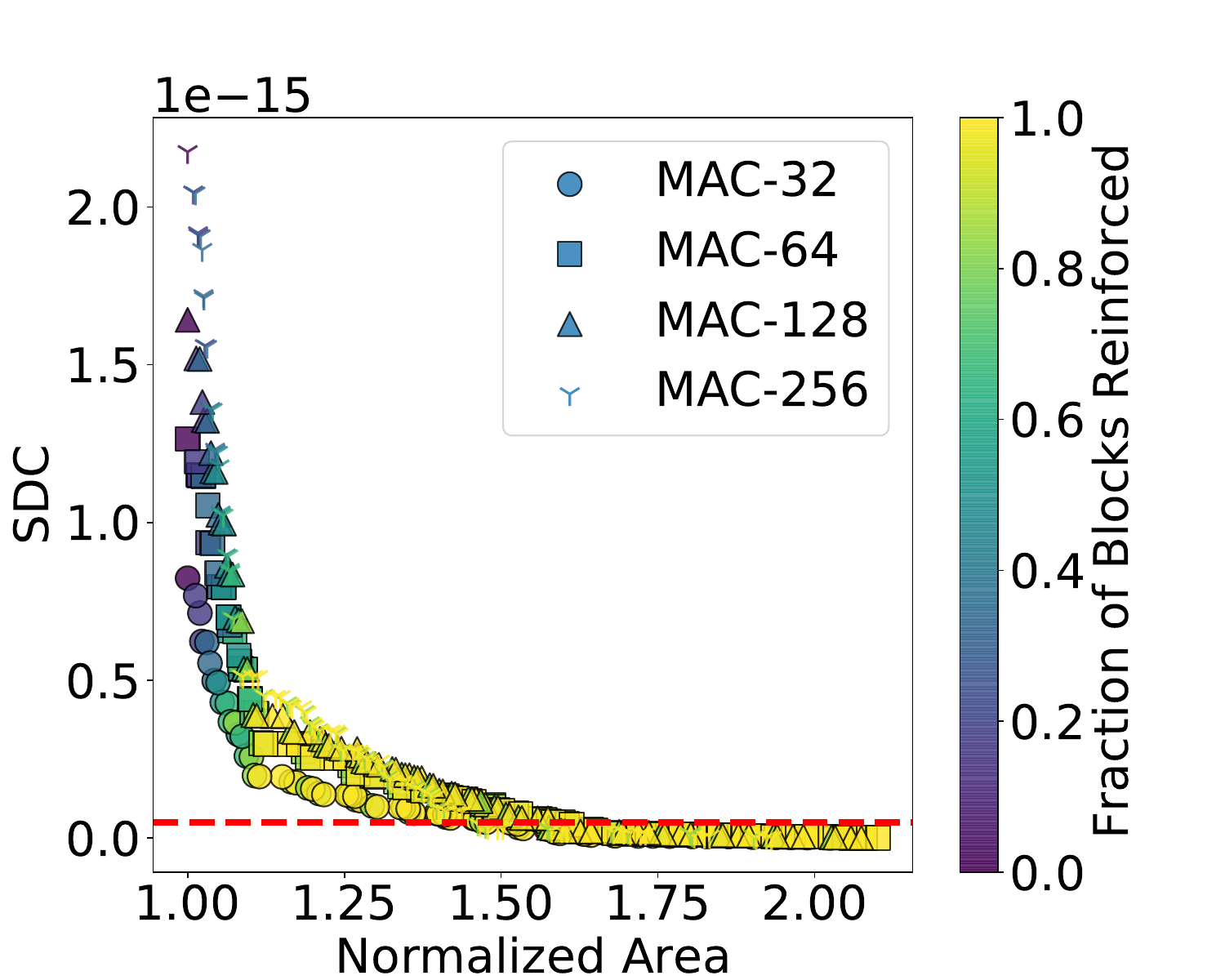}
		\label{fig:u55_mix}
	}
	\hspace{1pt}
	\caption{SDC rate per inference vs. area running Wav2Letter at TSMC 16nm and using a) DMR b) DMR with logic faults considered c) Flop hardening d) Flop hardening with logic faults considered e) Flop hardening supporting logic fault elimination and f) using a mixture of DMR, flop hardening, and flop hardening supporting logic fault elimination for the functional blocks in Ethos-U55.}
	\vspace{-10pt}
	\label{fig:fig:u55dmr}
\end{figure*}
With the analytical model developed in~\Sect{sec:fblock}, we can now study the impact of conventional protection schemes on U55's resiliency.
We start by discussing our experimental setup (Sec~\ref{sec:exp_set}) for estimating the resiliency of possible Ethos-U55 configurations.
We then show how DMR, flop hardening, and a mix of the two techniques impact the overall resiliency of Ethos-U55 (Sec~\ref{sec:study_res}).

\subsection{Experimental Setup}
\label{sec:exp_set}


%
%
\subsubsection{Resilient Configurations}
\label{sec:res_config}
The formulation in Equ~\ref{eq:sdcnpu_block_samp} is general; the component-wise SDC ($SDC_i$) and faulty probability ($P_1$), however, change with the protection scheme listed in~\Tbl{table:schemes}, which we describe next.

When \textbf{DMR} is used, it is assumed that none of the errors will go undetected from the block. This results in

\begin{align}
SDC_{Block} = 0,
\end{align}

\noindent and can be used in Equ~\ref{eq:sdcnpu_block_samp} to estimate $SDC_{NPU}$.
	\begin{table}[t]
		\centering
		\caption{Resilience techniques and associated parameters for estimating $SDC_{NPU}$. $FIT_{Hardened}$ is the FIT rate of the hardened FF and $\delta$ is area overhead of the checking logic.}
		\label{table:schemes}
		\resizebox{\columnwidth}{!}{%
			\begin{tabular}{ccccc}
				\toprule[0.15em]
			Type &Technique & \specialcell{Area\\ Cost} &  $\frac{FIT_{Hardened}}{FIT_{Unhardened}}$ & $\frac{\alpha^{'}}{\alpha}$\\
				\midrule[0.05em]
				Redundancy & DMR & 100 + $\delta$\%                                      & - & 0\\ 
				FF Hardening & Quatro~\cite{jahinuzzaman2009soft}                      & 157\%                            &    0.98 & 1\\ 
				FF Hardening & TSPC-DICE~\cite{jahinuzzaman2010tspc} &46.05\%                                          &              0.75 & 0\\ 
				\bottomrule[0.15em]
			\end{tabular}
		}
	\end{table}

Similarly, when \textbf{FF hardening} is used for flops in a functional block, it reduces the probability of a bit-flip occurring in any FF present in that block as hardening results in a reduction in the raw FIT rate of the FF. This implies that

\begin{align}
	P_{Block}^{'} = \frac{FIT_{Hardened}}{FIT_{Unhardened}}\times P_{Block}
\end{align}

\noindent where the ratio of the raw FIT rate of a hardened flop to that of an unhardened flop is listed in~\Tbl{table:schemes}. The calculated $P_{Block}^{'}$ becomes the probability of the block under FF hardening and can replaces $P_{Block}$ in Equ~\ref{eq:sdcnpu_block_samp} to estimate $SDC_{NPU}$.

Lastly, when FF hardening with SET elimination is used, it reduces the probability of a bit-flip occurring in the flop, and we can use Equ~\ref{eq:prob_logic_alpha} to estimate the new fault probability as

\begin{align}
		P_{Block}^{'} = \frac{FIT_{Hardened} + \alpha^{'}}{FIT_{Unhardened} + \alpha} \times P_{Block}
\end{align}

\noindent where $\alpha^{'}$ is the new increase in the raw FIT rate of the FF.
The calculated $P_{Block}^{'}$ becomes the probability of the block under FF hardening and can be used in Equ~\ref{eq:sdcnpu_block_samp} to estimate $SDC_{NPU}$.\\

\subsubsection{Area Evaluation}
\label{sec:area_eval}
We use the Synopsys Design Compiler with the TSMC 16nm and 7nm library to obtain the area numbers for the Arm Ethos-U55 base configurations of MAC-32, MAC-64, MAC-128, and MAC-256. We estimate the area overhead of each protection scheme to each configuration in~\Tbl{table:schemes}.
Specifically, we synthesized the checking logic, which is estimated to have an overhead ($\delta$ in~\Tbl{table:schemes}) of 7.3\%, 8.4\%, 10.7\% and 13.5\% at 16 nm and 5.1\%, 6.6\%, 7.4\% and 10.1\% at 7 nm for MAC-32, MAC-64, MAC-128 and MAC-256 respectively.


\subsection{Results}
\label{sec:study_res}
\subsubsection{\textbf{Ethos-U55 with functional block DMR}}
With functional block level DMR, Ethos-U55 is able to reduce its SDC rate down to ASIL-D levels with around 2$\times$ area overhead for all the MAC configurations.
~\Fig{fig:u55_dmr} shows SDC rate per inference vs. area of various configurations of Ethos-U55. Each block in U55 can either be left unprotected or be protected by either DMR, flop hardening, or flop hardening supporting logic fault elimination. The heatmap shows out of 6 functional blocks in Ethos-U55, what fraction of blocks are protected. We show only the Pareto optimal configurations. The horizontal dashed line shows the required SDC rate per inference to meet ASIL-D standards as calculated in Sec III-B.

If we look at the bottom right of~\Fig{fig:u55_dmr}, we find that configurations yielding the lowest SDC rates have almost all the functional blocks redundant because with DMR we either make an entire block redundant, or we do not, latter resulting in unprotected blocks contributing to the overall $SDC_{NPU}$. 

DMR can also detect soft-errors that might occur due to faults in combinational logic. And that is why, for the optimal configurations in~\Fig{fig:u55_dmr_logic} no extra silicon is spent as compared to the configurations in~\Fig{fig:u55_dmr} to achieve the lowest possible SDC rate when logic faults are considered. 

\subsubsection{\textbf{Ethos-U55 with flop hardening}}
\label{sec:u55_hard}

Ethos-U55 is not able to achieve the required SDC rate to meet ASIL-D standard when block-level flop hardening is employed as shown in Fig~\ref{fig:u55_hard}. We see that with just around 60\% area overhead, flop hardening $SDC_{NPU}$ gets extremely close to the desired SDC rate when all the blocks are hardened. We can understand this behavior by looking at the Equ~\ref{eq:sdcnpu_block_samp}, where we see that $SDC_{NPU}$ is dependent on the individual SDCs of the functional block, along with the probability of a soft-error occurring in that block. When flops are hardened in a block, it reduces the probability of a soft-error occurring in the block (in this case by 98\%), but is not sufficient to achieve the desired NPU level SDC rate.

Fig~\ref{fig:u55_hard_logic} shows that just flop hardening is also not enough to achieve ASIL-D level SDC rate when logic faults are taken into consideration. We use Quatro~\cite{jahinuzzaman2009soft} FFs for our analysis which do not offer protection against SETs, i.e. if a SET carrying enough charge, travels to the input of a hardened flop while meeting the setup and hold timing requirements, the FF will capture it as a normal input resulting in a bit flip due to a combinational logic error. 

\subsubsection{\textbf{Ethos-U55 with flop hardening and SET protection}}
When TSPC-DICE~\cite{jahinuzzaman2010tspc} FFs are used for mitigating soft-errors in Ethos-U55, we observe that U55 does not meet the ASIL-D level SDC rate for any of the MAC configurations as shown in Fig~\ref{fig:u55_hard_set}. Moreover, when compared with the resiliency of Etho-U55 with Qautro~\cite{jahinuzzaman2009soft} FFs we see that Quatro FFs overall achieve a SDC rate much closer to the desired levels as compared to the TSPS-DICE ones. This is because even though TSPC-DICE FFs can mitigate SETs (and hence logic errors), it does not reduce the probability of a fault occurring at a fault site to the same extent as a Quatro FF which results in a poor resiliency performance.


\subsubsection{\textbf{Ethos-U55 with a mix of DMR, flop hardening, and SET protection}}
 In this evaluation, each block can either be left unprotected or use one of either DMR, Quatro FFs, or TSPSC-DICE FFs.~\Fig{fig:u55_mix} shows that while using a combination of DMR, Quatro FFs, and TSPC-DICE FFs, we can achieve our required resiliency with as low as 53\% increase in the silicon. This is a significant improvement upon the configurations that used DMR, Quatro FFs, and TSPC-DICE FFs in isolation. Also, as evident from~\Fig{fig:u55_mix}, there are multiple different configurations available that have the required level of resiliency, giving designers the option to choose from to optimise for power and performance as well. 

We see from ~\Fig{fig:u55_mix} that there is a sharp decrease in the SDC rate as the functional blocks are protected. With just 15\% area overhead, we are able to achieve an SDC rate of around $0.3 \times 10^-15$, but to reach the desired ASIL-D standard SDC rates, another 30\% silicon area is required. We discuss the reasons for this behavior along with our findings from the optimal configurations of~\Fig{fig:u55_mix} in Sec~\ref{sec:res:find}.
\subsubsection{\textbf{What did we learn about Ethos-U55?}}
\label{sec:res:find}
\begin{table}[t]
	\centering
	\caption{U55 configurations meeting the ASIL-D SDC rate per inference requirement at TSMC 16nm and 7nm technology nodes. Here, 0 = No Protection, 1 = FF Hardening, 2 = DMR, and 3 = FF Hardening supporting logic fault elimination. Block Order: [AO, DMA, MAC, REG, TSU, WD].}
	\renewcommand*{\arraystretch}{1}
	\renewcommand*{\tabcolsep}{2pt}
	\resizebox{0.7\columnwidth}{!}
	{
		\begin{tabular}{ccc}
			\toprule[0.15em]
			 MAC Config & 16 nm  & 7 nm \\
			\midrule[0.05em]
			MAC-32       & [1, 3, 3, 3, 2, 2]  & [3, 3, 3, 3, 2, 2]  \\
			MAC-64       & [1, 3, 3, 3, 2, 2]  & [1, 3, 3, 3, 2, 2]   \\ 
			MAC-128 & [2, 3, 3, 3, 2, 2] & [2, 1, 3, 3, 2, 2]  \\
			MAC-256 & [2, 2, 2, 3, 2, 2]  & [2, 2, 2, 3, 2, 2]  \\
			\bottomrule[0.15em]
		\end{tabular}
	}
	\label{tbl:opt_config}
\end{table}

~\Tbl{tbl:opt_config} lists the Ethos-U55 configurations that achieve ASIL-D level resiliency under area constraints for all four MAC configurations. We see that for all 8 configurations TSU and WD functional blocks have DMR as the preferred technique for mitigating the effects of soft-error. This is expected because TSU and WD blocks have the largest block level SDC rate per inference as shown in~\Fig{fig:func_block} and DMR ensures that these blocks have zero contribution to the $SDC_{NPU}$ in the optimal configurations. We also observe that for MAC-32 and MAC-64 configurations, required resiliency can be achieved without duplicating half of the blocks and hence saving up on the silicon area. 

If we look at the optimal configurations for both 16 nm and 7 nm, we see that the blocks that occupy the highest area (DMA in this case) are avoided for both DMR and flop hardening as both techniques have a huge area overhead, except for the case of MAC-256 configuration.  In the case of MAC-256, most blocks are made redundant as the SDC contribution of each individual functional block is highest for MAC-256 among all the MAC configurations.

We see that the optimal configurations have similar structures for both 16nm and 7nm technology nodes for all the MAC configurations. However, the overall area overhead of the optimal configuration in 7nm is 21.7\% less than that of the same configuration in 16nm owing to the reduction in silicon area due to technology scaling.

\section{Related Work}
\label{sec:related}
The main novelty of our work is three-fold. First, we carry out a large-scale, RTL-based, reliability analysis of a commercial NPU that is currently used by a number of customers in the market. 
Other than works on GPUs~\cite{dos2018analyzing,ibrahim2020analyzing,wei2020analyzing,ibrahim2020soft}, most of the reliability analysis is carried out by making use of non-commercial ML inference accelerators. G. Li et al.~\cite{li2017understanding} use accelerators such as Diannao~\cite{chen2014diannao} DaDiannao~\cite{chen2014dadiannao}, and Eyeriss~\cite{chen2016eyeriss}. Reagan et al.~\cite{reagen2018ares} carries out reliability analysis on their in-house accelerator~\cite{whatmough201714}, and so do Choi et al.~\cite{choi2019sensitivity} and Zhang et al.~\cite{zhang2018thundervolt}.

Commercial accelerators such as NVDLA~\cite{nvdla}
have been characterized for their soft-error reliability with Fidelity~\cite{he2020fidelity}, and TPUs~\cite{jouppi2017datacenter}have been analyzed by Rech et al.~\cite{rech2022reliability}. Rech et al. carry out beam experiments to study the reliability, which involves a sophisticated and not readily accessible facility and Fidelity makes use of a framework validated on 15X fewer fault injections (larger the number of fault injections, the higher the accuracy) as compared to our work.

Secondly, we characterize the functional blocks of the NPU for their soft-error resiliency behavior as the functional blocks are common across various designs of ML inference accelerators. Whereas, prior works have primarily focused on two factors related to accelerators: memory~\cite{ibrahim2020analyzing,lee2014fault,choi2019sensitivity,clemente2016hardware}, and processing element~\cite{ibrahim2020analyzing,jiao2017assessment,choi2019sensitivity}. 

Lastly, we analyze the effects of heterogeneous soft-error protection schemes, where one selectively applies different protection strategies to different functional blocks.
Selective resiliency methods are not new and have been studied at the architecture level~\cite{nikseresht2021selective,feng2010shoestring}, application level~\cite{mahmoud2021optimizing,hanif2020dependable,wan2023vpp} and hardware level~\cite{huang2020functional,hanif2020dependable}. To that end, our paper presents detailed studies in soft-error resiliency of individual functional blocks, which are missing in the prior art and can be useful for future studies.
\section{Conclusion}

We perform a thorough characterization of the Arm Ethos-U55 NPU, which targets embedded space, against soft errors. We show that while U55 is designed to meet ASIL B/C standards, it does not meet the ASIL D standard.
In order to meet the ASIL D standard, we should that a calculated trade-off between area and resiliency must be made.
We show that selectively duplicating certain function blocks while hardening FFs in others allows us to meet the ASIL D standard while minimizing the area overhead.

\section{Acknowledge}

We thank anynomous reviewers from ISPASS 2024 for their valuable comments.
We thank the Arm Academic Access program for providing us accesses to the Ethos-U55 IP and the associated tools.
The research is partially supported by NSF Award \#2044963 and a gift grant from Arm.



\bibliographystyle{IEEEtranS}
\bibliography{refs}
\end{document}